\documentclass[11pt]{article}

\usepackage[T1]{fontenc}
\usepackage[utf8]{inputenc}
\usepackage{amssymb}
\usepackage{amsmath}
\usepackage{lineno}
\usepackage{booktabs}
\usepackage{tabularx}
\usepackage{graphicx}
\usepackage{multirow}
\usepackage{array}
\usepackage{longtable}
\usepackage{xcolor}
\usepackage{hyperref}
\hypersetup{
  colorlinks=true,
  linkcolor={blue!70!black},
  citecolor={blue!70!black},
  urlcolor={blue!70!black}
}
\usepackage{url}
\usepackage{doi}
\makeatletter
\g@addto@macro{\UrlBreaks}{\UrlOrds}
\makeatother
\usepackage[margin=1in]{geometry}
\usepackage{float}
\usepackage{placeins}
\usepackage{caption}
\usepackage{subcaption}
\usepackage{threeparttable}
\usepackage{makecell}
\usepackage{newtxtext}
\usepackage{newtxmath}
\usepackage{setspace}
\usepackage{titlesec}
\usepackage[numbers,sort&compress]{natbib}
\usepackage{authblk}

\renewcommand{\arraystretch}{1.08}
\titleformat{\section}{\normalfont\Large\bfseries}{\thesection.}{0.65em}{}
\titleformat{\subsection}{\normalfont\large\bfseries}{\thesubsection}{0.6em}{}
\titleformat{\subsubsection}{\normalfont\normalsize\bfseries}{\thesubsubsection}{0.6em}{}
\titleformat{\paragraph}[runin]{\normalfont\normalsize\bfseries\itshape}{}{0pt}{}
\titlespacing*{\section}{0pt}{1.8ex}{0.7ex}
\titlespacing*{\subsection}{0pt}{1.4ex}{0.5ex}
\titlespacing*{\subsubsection}{0pt}{1.1ex}{0.4ex}
\titlespacing*{\paragraph}{0pt}{0.6ex}{0.5em}
\graphicspath{{Figures/}}

\begin{document}

\title{\LARGE\bfseries AI Data Centers and the Water Use Feedback Loop}

\author[1]{Basit A. Akinade}
\author[1]{Amobichukwu C. Amanambu\thanks{Corresponding author. Email: acamanambu@ua.edu}}
\author[2]{Jonathan M. Frame}
\author[3]{Shaolei Ren}
\affil[1]{Water INtelligence and Geospatial Sensing (WINGS) Laboratory, Department of Geography and the Environment, The University of Alabama, Tuscaloosa, AL, USA}
\affil[2]{Department of Geological Sciences, The University of Alabama, Tuscaloosa, AL, USA}
\affil[3]{Department of Electrical and Computer Engineering, University of California, Riverside, CA, USA}

\date{}
\maketitle

\begin{abstract}
AI data centres consume water for cooling, water scarcity constrains siting, and AI tools can improve water system efficiency. These dynamics are studied separately yet form a feedback loop. This review formalises the Water and AI Feedback Loop, introduces the Water Consumption Impact index to quantify community-scale utility burden, and demonstrates across ten US sites that burden spans three orders of magnitude, from 0.2\% to 134\% of host capacity.
\end{abstract}


\section{Introduction}
\label{sec:intro}

Artificial intelligence (AI) is transforming global economies faster than society can evaluate its environmental footprint. Data centres already account for roughly 4--5\% of U.S. electricity consumption at 177--192~TWh in 2024 \citep{Shehabi2024}, projected to reach 380--790~TWh by 2030 \citep{EPRI2026}. This growth dwarfs the 1.2--3.3\% annual increase in total U.S. electricity demand forecast for 2026--2027 \citep{EIA_STEO2026}, making data centres the dominant driver of marginal grid load in many regions. Each kilowatt-hour of electricity consumed by AI workloads incurs water costs across three dimensions: direct cooling water evaporated on site, indirect water consumed at power plants, and supply-chain water embedded in semiconductor fabrication \citep{Mytton2021,Siddik2021}. Training GPT-3 alone consumed roughly 700,000 litres of freshwater for on-site cooling and 5.4 million litres once electricity-related use is included \citep{Li2025}. Inference is now scaling faster than training \citep{Shehabi2024}, and per-query water footprints vary from under 4~mL to more than 200~mL across architectures \citep{Jegham2025}. Yet these three dimensions have been studied in isolation \citep{Lei2025,deVriesGao2026,WaterAINexus2025,BarnettItzhaki2026}. 

Water infrastructure is an \textit{active constraint} that governs where data centres can be built, which cooling technologies they may adopt, and how quickly they can expand. In Uruguay, Google reformulated a planned hyperscale facility, ultimately adopting air cooling in place of evaporative systems, after the country's worst drought in seven decades and public opposition over its projected 7.6~ML/d potable demand \citep{DCD_Uruguay2023,Mongabay_Uruguay2023}. In Chile, an environmental court partially overturned Google's permit in 2024 \citep{Tironi2025}. The Netherlands and Ireland have imposed moratoriums on new construction citing cumulative water and energy pressure \citep{DCD_Netherlands2022,CRU_Ireland2025}. The City of Lebanon, Indiana is undertaking a phased infrastructure expansion projected to take six years before a Meta campus can reach full water capacity \citep{CityLebanon2025,Han2026}, and industry analysts increasingly identify water availability, not power, as the emerging binding constraint on siting \citep{SPGlobal2025,Han2026}. These are not isolated incidents but a systemic global pattern.

AI itself is being deployed \textit{inside} water systems. Machine learning is being deployed to detect leaks in ageing distribution networks \citep{Hayslep2025,Taloma2025}. Also, AI is increasingly being used for forecasts, including demand for capacity planning \citep{Fu2022}, simulations of infrastructure stress through digital twins \citep{WaterTwinAI2025}, and predictions of streamflow for drought early warning \citep{Kratzert2019,Nearing2021}. Globally, utilities lose roughly 30\% of treated water as non-revenue water or through leaks and other system losses \citep{LiembergerWyatt2019}, a volume that dwarfs total data centre use. Yet no framework connects the AI-for-water problem to the water-for-AI problem.

The AI and water relationship is a loop, not a line: three pathways interact in a feedback structure, each rooted in a distinct disciplinary domain. \textit{Burden} pathways (hydrology), through which data centres stress source watersheds and municipal systems. \textit{Constraint} pathways (infrastructure engineering), through which pipe capacity, treatment limits, and regulatory conditions reshape AI development. \textit{Adaptive} pathways (computer science and AI), through which machine learning tools can recover capacity and reduce net stress (Figure~\ref{fig:triad}). This structure is the Water and AI Feedback Loop.

This paper argues that the AI and water problem is fundamentally a \textit{community-scale utility burden problem}, not a national volumetric one. Aggregate contributions appear modest nationally, but this masks local realities in which a single facility's peak consumptive demand can press against 30\% or more of a host utility's service capacity \citep{Han2026} and create burden intensities that vary by more than three orders of magnitude, depending on three site-specific factors: facility size relative to the utility, the fraction of water lost to evaporation, and the severity of peak-day spikes. We demonstrate the framework using U.S. sites where standardised utility capacity data are most available, while designing the tools to be transferable to any jurisdiction. To anchor this assessment, we introduce the Water Consumption Impact (WCI) index, defined as the fraction of a host utility's peak-day delivery capacity consumed by a single facility's evaporative cooling demand. The index is developed and applied in Section~\ref{sec:synthesis}.

Whether the loop reinforces burden or enables adaptation depends on whether it is intentionally managed. Unmanaged, it is self-reinforcing: rising water stress forces less energy-efficient cooling, which raises peak cooling power demand and indirect water consumption and leaves less power capacity for compute, concentrating facilities in progressively less suitable locations. Managed, it can become self-correcting: AI tools improve water efficiency, freed capacity enables energy-efficient cooling, and reduced cooling power demand decreases indirect water use and increases power capacity for compute. No governance framework or corporate strategy identified to date treats the three pathways as a coupled physical system. The paper proceeds from framework to burden, constraint, and adaptation, then to integrated synthesis and governance priorities.

\section{Review Scope and Conceptual Framework}
\label{sec:framework}

\subsection{Framework Conceptualization from Fragmented Literature}
\label{sec:framework_frag}

The literature supports three related claims, but usually in isolation: (1) AI water use burdens source watersheds and host communities \citep{Siddik2021,Han2026}, (2) water systems constrain where AI grows \citep{Han2026,SPGlobal2025}, and (3) AI tools can improve water-system performance \citep{Taloma2025,Fu2022}. What is missing is a single framework connecting these pathways within real host systems. Table~\ref{tab:silos} summarises the disciplinary split, and Table~\ref{tab:triad_synthesis} condenses the core logic. Engineering and computer science studies quantify Water Usage Effectiveness (WUE) and volumetric footprints but ignore source hydrology and pipe capacity \citep{Lei2022,Mytton2021,Li2025,Lei2025}. Water resources studies resolve spatial footprints and streamflow stress but miss the infrastructure bottlenecks that determine whether water actually reaches a facility \citep{Siddik2021,Siddik2024,Amanambu_HSI,ICPRB2025}. Infrastructure analyses document peaking factors and municipal capacity limits but miss hydrology and AI-enabled adaptation \citep{Han2026,WaterAINexus2025}. Identical facility demand is therefore treated as comparable across locations when in practice its significance depends on source hydrology, municipal capacity, and the possibility of local recovery. \citet{Lei2025} note the absence of hydrological analysis in workload-level WUE studies, and \citet{deVriesGao2026} remain at corporate-aggregate volumetrics.

\begin{table}[H]
\centering
\caption{Three disciplinary silos in the AI and water literature. Each silo addresses a subset of the coupled system but misses the remaining pathways.}
\label{tab:silos}
\small
\begin{threeparttable}
\begin{tabularx}{\textwidth}{>{\raggedright\arraybackslash}p{2.8cm} >{\raggedright\arraybackslash}p{3.5cm} >{\raggedright\arraybackslash}X >{\raggedright\arraybackslash}X}
\toprule
\textbf{Silo} & \textbf{Representative works} & \textbf{Coverage} & \textbf{Gaps} \\
\midrule
Engineering / Computer Science & \citet{Lei2022}, \citet{Mytton2021}, \citet{Li2025}, \citet{Lei2025}, \citet{Karimi2022} & WUE, cooling technology, volumetric accounting, per-query footprints & Hydrology, infrastructure capacity, water's constraint on AI \\[6pt]
Water Resources & \citet{Siddik2021}, \citet{Siddik2024}, \citet{Amanambu_HSI}, \citet{ICPRB2025} & Spatial footprints, streamflow stress, watershed impacts & Infrastructure capacity, AI tools for water, water's constraint on AI \\[6pt]
Infrastructure / Policy & \citet{Han2026}, \citet{WaterAINexus2025} & Pipe capacity, peaking factors, public water system constraints & Hydrology, AI for water management \\
\bottomrule
\end{tabularx}
\end{threeparttable}
\end{table}

\subsection{A Coupled Framework for the AI and Water Nexus}
\label{sec:framework_coupled}

\begin{figure}[H]
\centering
\includegraphics[width=0.72\textwidth]{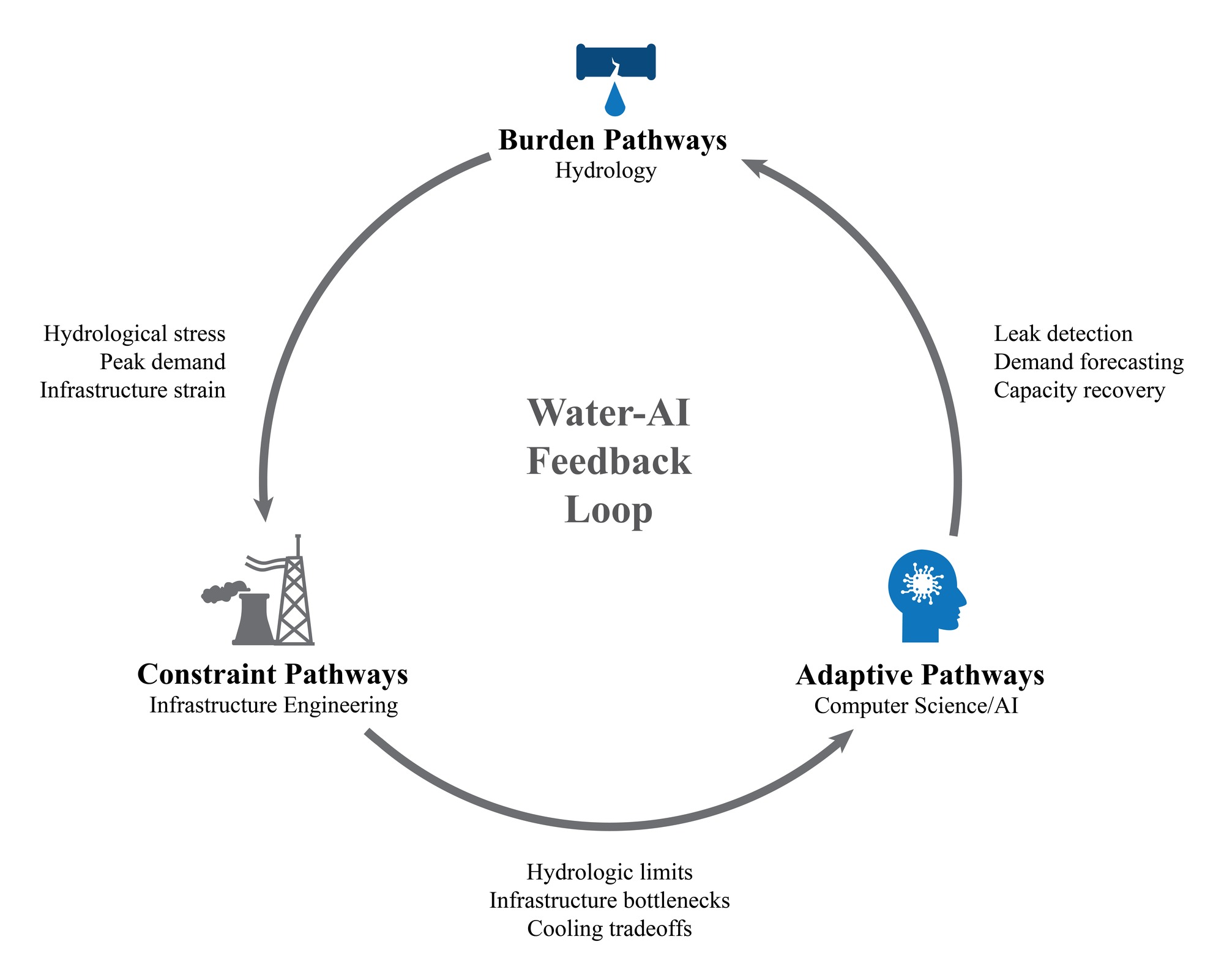}
\caption{The Water and AI Feedback Loop. Three coupled pathways, each rooted in a distinct disciplinary domain, form a feedback structure: burden pathways (hydrology) through which AI stresses source watersheds and municipal systems, constraint pathways (infrastructure engineering) through which pipe capacity and treatment limits reshape AI development, and adaptive pathways (computer science) through which AI tools can conditionally recover water system capacity. The loop's trajectory depends on whether the coupling is intentionally managed.}
\label{fig:triad}
\end{figure}

The relationship between AI infrastructure and water systems is tripartite, with three coupled pathways forming a feedback loop (Figure~\ref{fig:triad}).

\textit{Burden pathways} are the mechanisms through which data centres impose hydrological, infrastructure, and water-quality stress on the systems they draw from. These burdens are nationally modest but can be locally severe, because data centres combine extreme-->high consumptive ratios (70--90\% versus a 12\% public supply average) with extreme-->high peaking factors (6--30 versus 1.5--2.5 for other users) \citep{Han2026,Medalie2025}. These two properties interact with location-specific hydrology and infrastructure capacity to produce stress outcomes varying by more than four orders of magnitude across sites \citep{Amanambu_HSI}.

\textit{Constraint pathways} are the infrastructure mechanisms through which water systems limit or reshape the geographic distribution of AI infrastructure, architecture, economics, and/or timeline. This leads to potential capacity bottlenecks in treatment, transmission, and storage that cannot deliver water at data centre peak rates, and may require technology substitution. It also creates the need for regulatory intervention when capacity is insufficient.

\textit{Adaptive pathways} are the conditional potential, rather than demonstrated closure, of AI tools deployed within water systems to recover lost capacity and reduce net stress. This potential is conditional on local and temporal matching (do the savings occur in the same system and peak period as the burden?), recoverable volume, implementation cost, and governance feasibility.

\subsection{Evidence Base, Source Types, and Review Logic}
\label{sec:framework_methods}

In this analysis we draw on 97 sources identified through systematic searches of Web of Science, Scopus, and Google Scholar, supplemented by citation tracking and targeted grey literature searches.\footnote{The bibliography contains 104 entries; seven additional references (operator fact sheets, site-level data sources, and unit-conversion constants) serve as data inputs for the WCI analysis and are not counted in the systematic literature total.} Sources were included if they provided quantitative data on data centre water use or AI tool performance, documented cases of water constraining AI development, or offered analytical frameworks bridging two or more pathways. Four evidence tiers are distinguished, in descending weight: 1) peer-reviewed research, 2) government and regulatory documents, 3) corporate disclosures (treated as self-reported), and 4) investigative reporting (used for case context). Because peer-reviewed literature on water as a constraint on AI remains sparse, the constraint pathway draws on public records and grey literature, ranked by evidentiary strength. Supplementary Tables~S1, S3 and~S4 audit constraint sources, classify all 97 sources by pathway and tier, and summarise the totals. The analysis focuses on water quantity and infrastructure capacity, with water quality addressed briefly.

\subsection{Cross-Scale Feedbacks, Temporal Mismatch, and Leverage Points}
\label{sec:framework_feedbacks}

Three properties of the coupled system explain why the loop can produce either reinforcing burden or adaptive benefit, and cut across all three pathways.

\textbf{Location dominance.} Hydrological sustainability depends mostly on where a facility is sited, not on its cooling technology: the location effect on stress is approximately 4{,}897 times larger than the technology effect \citep{Amanambu_HSI}. Maximum technology improvement cannot compensate for a hydrologically unfavourable location, affecting both the constraint pathway (water-scarce sites force substitution) and the adaptive pathway (savings matter only where the burden falls).

\textbf{Temporal synchrony.} Peak cooling demand, low streamflow, high grid stress, and peak community demand coincide in hot climates, a thermodynamically inevitable quadruple coincidence. The Drought Amplification Factor formalises this, ranging from near unity on large regulated rivers to over 100 in flashy desert systems \citep{Amanambu_HSI}. The binding constraint is not the annual average but the worst hour of the driest year, precisely when cooling demand peaks. Any adaptive pathway that does not address this temporal dimension fails to close the loop.

\textbf{Peak demand as binding constraint.} For both rivers and pipes, the binding constraint is peak demand, not annual volume. A nationally modest 0.6--1.1\% contribution to public withdrawals by 2030 \citep{Han2026} masks local peak-day stress that can exceed infrastructure design capacity by an order of magnitude; data centre peaking factors of 6 to over 30 \citep{Han2026,ICPRB2025} mean infrastructure designed for average demand fails under peak load.fails under peak load. Even where reservoir storage ensures adequate raw water supply, the treatment plant and distribution network remain rate-limited; storage volume does not increase the maximum daily throughput of the delivery system. Peak stress directly triggers the bottlenecks, opposition, and regulatory responses that constitute the constraint pathway.

\subsection{The Water Blind Spots in Current Evidence and Governance}
\label{sec:framework_blindspots}

The most consequential pressures emerge where facility demand, watershed stress, and municipal infrastructure limits intersect, yet these interactions are rarely evaluated together. Five blind spots explain why local capacity can fail even when annual water use appears manageable:

\begin{enumerate}
    \item \textbf{Measurement.} No dataset links facility-level cooling demand with real-time streamflow, groundwater, municipal capacity, and peak indicators. The ( Hydrological Stress Index) HSI framework \citep{Amanambu_HSI} and pipe-capacity analysis of \citet{Han2026} each address one dimension, but utility Supervisory Control and Data Acquisition (SCADA) and data centre cooling controllers operate in separate data silos.

    \item \textbf{Cumulative impact.} Assessments evaluate facilities individually, not the cumulative burden of co-located data centres on shared watersheds, aquifers, or municipal systems. Phoenix already exceeds sustainable thresholds under cumulative demand \citep{Amanambu_HSI} and Northern Virginia's aggregate data centre peaking factor reaches 10, with consumptive use rising from 9\% to 12\% of the regional total during summer \citep{ICPRB2025,ICPRB2026}, yet permits treat each project independently.

    \item \textbf{Governance.} No decision framework jointly evaluates facility demand, hydrologic stress, infrastructure capacity, and offset potential. As of 2025, no jurisdiction has established a water permitting category specific to data centres \citep{Han2026}, and basin commissions operate independently of municipal planners \citep{SRBC2025}.

    \item \textbf{Offset evidence.} AI leak detection, digital twins, and demand forecasting show 95--99\% accuracy \citep{ZunigaUribe2026} and pilot-scale 15\% loss reductions \citep{WaterTwinAI2025}, yet no peer-reviewed evidence demonstrates that these interventions offset co-located data centre burden at operational scale.

    \item \textbf{Spatial mismatch.} Corporate Water Positive claims are reported at broad geographic scales, obscuring whether benefits occur in the same places, seasons, and systems where burdens fall \citep{Han2026,MetaLimnoTech2025}.
\end{enumerate}

\begin{table}[H]
\centering
\caption{Compact synthesis of the three pathways used throughout Sections~\ref{sec:burden}--\ref{sec:adaptive}. The table is a logic map rather than a literature inventory.}
\label{tab:triad_synthesis}
\footnotesize
\begin{threeparttable}
\begin{tabularx}{\textwidth}{>{\raggedright\arraybackslash}p{1.7cm} >{\raggedright\arraybackslash}p{2.4cm} >{\raggedright\arraybackslash}X >{\raggedright\arraybackslash}p{2.7cm} >{\raggedright\arraybackslash}X >{\raggedright\arraybackslash}X}
\toprule
\textbf{Pathway} & \textbf{Core mechanism} & \textbf{What is already established} & \textbf{Anchor indicator / evidence base} & \textbf{Main unresolved issue} & \textbf{Implication for this framework} \\
\midrule
Burden & Data centres impose stress through watershed withdrawals, peak utility demand, and cooling tradeoffs. & Water use is locally concentrated, highly consumptive, and often amplified during hot, dry periods; location and peak demand matter more than annual totals alone. & Hydrological stress analysis \citep{Amanambu_HSI}; peak-capacity and peaking-factor analysis \citep{Han2026}; WUE and facility cooling studies \citep{Lei2022,Karimi2022}. & How to evaluate burden jointly across source hydrology, host utility capacity, and compound local stressors. & Burden must be treated as both a watershed problem and a host-utility problem. \\
\addlinespace
Constraint & Water systems reshape AI growth through capacity bottlenecks, technology forcing, and regulatory access conditions. & Public systems can lack the peak deliverable capacity needed for hyperscale cooling, and these limits can alter siting, timing, and technology choice. & Permit filings, utility records, and public infrastructure cases \citep{Han2026,SRBC2025}. & No common decision framework yet links facility demand, source stress, infrastructure limits, and offset potential in one permitting logic. & Constraint is not downstream context; it is one side of the feedback loop and must be analysed alongside burden. \\
\addlinespace
Adaptive & AI tools can recover water-system capacity or reduce operational stress, but only under specific local conditions. & Leak detection, forecasting, digital twins, streamflow prediction, and reclaimed-water optimisation show technical promise. & Water-system AI reviews and operational pilots \citep{Taloma2025,ZunigaUribe2026,WaterTwinAI2025,Kratzert2019,Nearing2021}. & Whether improvements occur in the same place, season, and system where AI infrastructure imposes burden at operational scale. & Adaptation should be treated as conditional capacity recovery, not assumed offset, which is why the framework distinguishes reinforcing from enabling feedback regimes. \\
\bottomrule
\end{tabularx}
\end{threeparttable}
\end{table}

\section{Burden Pathways: How AI Loads Water Systems}
\label{sec:burden}

Table~\ref{tab:triad_synthesis} summarises the burden pathway. AI data centres impose water stress through on-site cooling evaporation (Scope~1), electricity-related consumption at power plants (Scope~2), and supply-chain water embedded in semiconductor fabrication and construction (Scope~3). Severity depends on where in the water cycle the stress falls and how concentrated it becomes in host communities. High consumptive ratios (70--90\%) mean most withdrawn water evaporates and may not return, reducing downstream flows and concentrating thermal and chemical loads \citep{Privette2026,Han2026}. Data centres may also potentially degrade host-community liveability through thermal effects \citep{Marinoni2026} and, emerging evidence suggests, property-value impacts near facilities \citep{Rubinovitz2026}.

\subsection{Source Watershed Stress}
\label{sec:burden_watershed}

At watershed scale, location dominates volume. The HSI framework evaluates withdrawals as a fraction of allocable drought-period streamflow across thirteen major U.S. hubs using 30-year USGS records \citep{Amanambu_HSI}. For a standardised 250~MW facility with a 30\% environmental reservation, drought-period HSI ranges from effectively zero at The Dalles, Oregon, to 42.78\% at Phoenix, Arizona, varying by more than four orders of magnitude; the maximum IT capacity that drought-period streamflow could in principle support spans 175~MW at Phoenix to a theoretical hydrologic ceiling exceeding 9{,}000~GW at The Dalles \citep{Amanambu_HSI}. The 9{,}000~GW figure is a water-only upper bound rather than a realistic build target, roughly seven times current total U.S. electricity generation capacity, and is best read as indicating that water is effectively non-binding at The Dalles while at Phoenix it constrains capacity at the scale of a single facility. The same 14~ML/d withdrawal registers as negligible stress (0.05\%) on the Missouri River but exceeds physical feasibility at 104\% on Arizona's Salt River and 152\% on Coyote Creek. Cumulative analysis reveals Phoenix has already exceeded its sustainable threshold (drought HSI 103\%) and San Antonio approaches infeasibility at 60\%.

Internationally, 72\% of Chinese data centre capacity sits in water-scarce regions \citep{Jiang2025}, and 43\% of data centres worldwide operate in high water-stress areas, projected to reach 45\% by the 2050s \citep{SPGlobal2025}, reflecting historical prioritisation of power, land, and fibre over water \citep{Siddik2021}. Ireland illustrates a distinct form of burden in which energy, not direct withdrawal, is the binding constraint: data centres accounted for 22\% of Ireland's metered electricity in 2024, projected at 32\% by 2026 \citep{CSO2025,CRU_Ireland2025}, translating into substantial indirect water consumption even with air-side cooling.

\begin{figure}[H]
\centering
\includegraphics[width=0.92\textwidth]{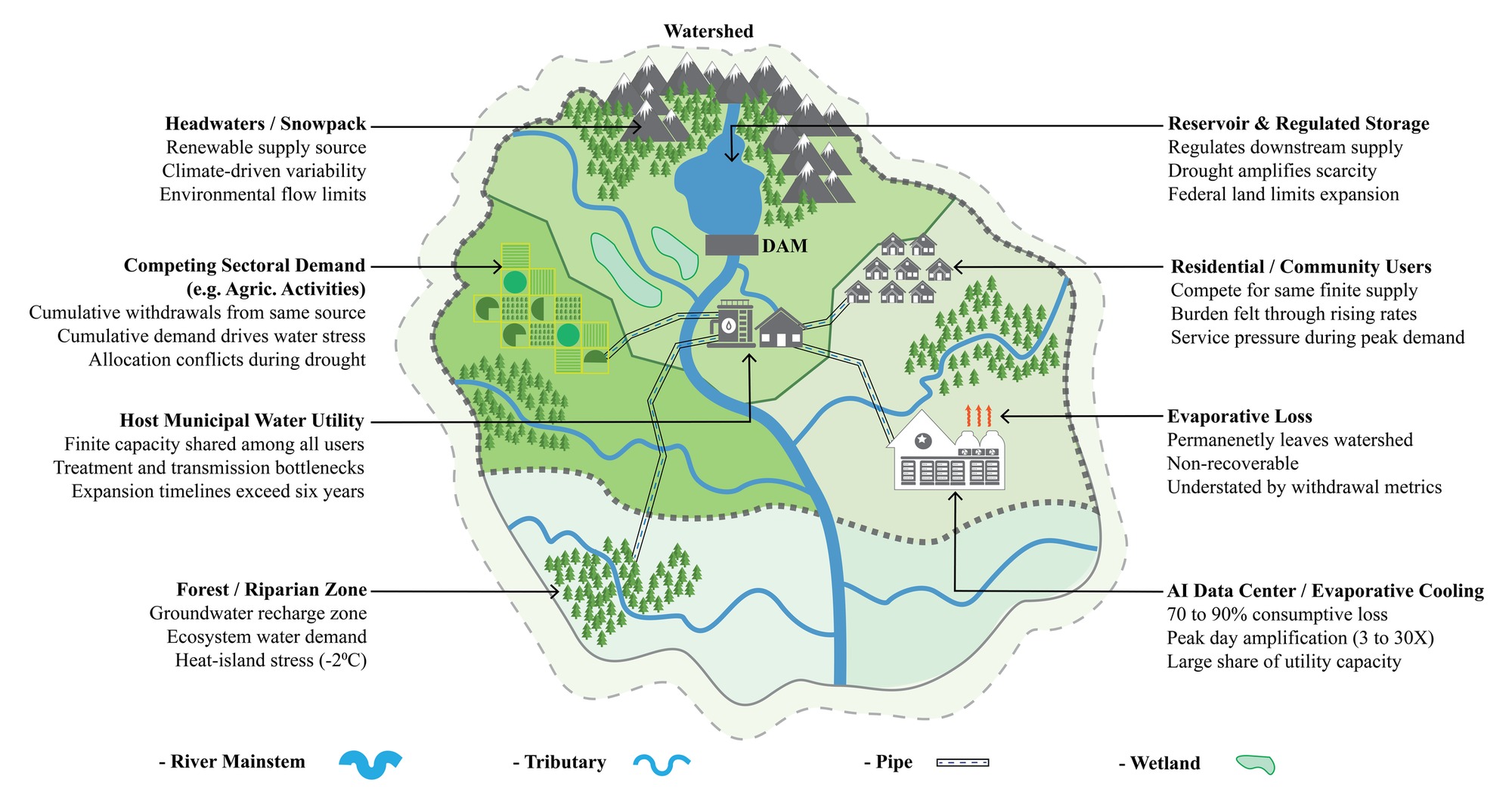}
\caption{Conceptual watershed accounting framework for AI data centre water consumption. The schematic traces water from headwater snowmelt and precipitation through regulated storage and municipal treatment to the major demand sectors within a host community. Evaporative cooling at the data centre consumptively removes 70--90\% of withdrawn water from the watershed, a consumptive loss not captured by conventional withdrawal statistics. The host utility's finite deliverable capacity is shared among residential, agricultural, and industrial users, and cumulative withdrawals during overlapping seasonal peaks can push the Hydrological Stress Index \citep{Amanambu_HSI} above unity. Adapted from \citet{Apple2025}.}
\label{fig:watershed_accounting}
\end{figure}

Within any single watershed, finite yield is shared among forests, irrigated agriculture, the host utility, and any co-located data centres, each with sharply different consumptive ratios and temporal profiles (Figure~\ref{fig:watershed_accounting}). Adding a data centre reallocates the shared yield rather than simply adding demand, with reallocation magnitude depending on where in the basin the facility is sited. The same facility imposes negligible stress on a high-yield basin and exceeds feasibility on a low-yield one, the central HSI observation and empirical foundation for the WCI analysis in Section~\ref{sec:wci_results}.

Hydrological stress is also a binding physical constraint on expansion. Phoenix has exceeded its sustainable threshold (HSI~$>$~100\%), making further growth impractical without inter-basin transfer or desalination. In The Dalles, continued expansion requires additional water from a federal reservoir, triggering environmental review and potentially federal legislative action \citep{OPB2026}. Watershed hydrology determines where AI data centres are and are not hydrologically suitable, forming the basis for the infrastructure constraints of Section~\ref{sec:constraint}.

\subsection{Municipal Infrastructure Stress}
\label{sec:burden_infrastructure}

Even where watersheds retain adequate capacity, the infrastructure delivering water may still fail at peak. The first comprehensive assessment of data centre impacts on U.S. public water systems shows the binding constraint is peak daily demand, not annual volume \citep{Han2026}. Two characteristics distinguish data centres from all other public water users. First, a high consumptive ratio: data centres evaporate 70--90\% of withdrawn water versus 5--15\% for residential users and a 12\% public supply average \citep{Han2026,Medalie2025}, disproportionately reducing downstream availability and complicating return-flow-based water rights. Second, an extreme peaking factor: 3 to over 30 for data centres versus 1.5--2.5 for typical public users \citep{Han2026}. West Des Moines hyperscale data show a monthly peak-to-average of 4.30, implying a daily peaking factor of at least 6.45 \citep{Han2026}. A Wisconsin Microsoft facility requested 0.7~MGD peak against average use of only 23{,}000 gallons, implying a peaking factor above 30 \citep{WisPubRadio2025,Han2026}. In Northern Virginia, the Prince William Water service area recorded a daily peaking factor of 10 \citep{Water_Planning_WashingtonMetro_2050Prediction_DataCenter_PeakingFactor_3point62_Report_2025}. These factors create bottlenecks at every component of the public water system.

Between 2024 and 2030, U.S. data centres could collectively require 697--1{,}451~MGD of new water capacity, comparable to New York City's average daily supply of $\sim$3,785~ML/d ($\sim$1{,}000~MGD), at an infrastructure cost of 7--58 billion USD \citep{Han2026}. These costs concentrate on individual host communities and often exceed the capacity of many community water systems already facing a 1.3--3.4 trillion USD national funding gap for water and wastewater \citep{EPA2023,ValueOfWater2025}.

\subsection{Facility-Level Consumption and the Water and Energy Tradeoff}
\label{sec:burden_facility}

At facility scale, cooling design translates heat load into both water and electricity demand, accounting for up to 45\% of total energy consumption \citep{Alkrush2024}. Climate- and technology-specific estimates across 15 U.S. climate zones show hyperscale WUE values of 0.09--1.80~L/kWh, while traditional evaporative systems require 1.99--4.00~L/kWh \citep{Lei2022}; the technology effect averages 10.57$\times$ across locations \citep{Amanambu_HSI}.

A fundamental tradeoff governs cooling design (Table~\ref{tab:cooling}). Evaporative cooling commonly reduces the peak cooling power demand by 25--35\% in the summer compared with dry cooling \citep{Han2026} but concentrates water use on the hottest days when water and grid stress are highest. The reduction in peak cooling power demand translates
into more power capacity available for computing under the same site-level power budget.
In hot-arid climates, Phoenix measurements show water-cooled chillers achieve lower energy but higher water use, and air-cooled systems the inverse \citep{Karimi2022}. Hybrid architectures offer compromise: switching to dry cooling in California reduced water use by 12.9~ML/month (3.4~Mgal/month) at a total cost increase of only 0.7\% \citep{FlexCoolDC2024}, and a 24-month NREL trial of a thermosyphon cooler hybrid reduced data centre water use by $\sim$50\%, lowering on-site WUE from 1.27 to 0.70~L/kWh without degrading energy performance \citep{Sickinger2018}.

Where systems discharge heated water, thermal pollution adds a downstream dimension: elevated temperatures impair aquatic ecosystems and trigger regulatory limits \citep{Privette2026}. Overlaying U.S. data centre footprints with freshwater species richness identifies data centres as an emerging biodiversity threat via thermal discharge, indirect water use at thermoelectric plants, and flow regime alteration at hydropower facilities \citep{JagerYoon2026}. 
Liquid cooling at the server rack reduces chip-level heat transfer to the facility cooling loop, but the facility must still reject that heat, often through evaporative systems that consume water \citep{RolandBerger2025}. Yet even state-of-the-art liquid-cooled facilities require substantial water for peak heat rejection: Meta's Indiana campus uses a closed-loop liquid-cooled system but still requires 30.3~ML/d (8~MGD) at full buildout \citep{Meta2026,CityLebanon2025}.

\begin{table}[H]
\centering
\caption{Cooling technology spectrum for data centers: water use intensity, energy penalty, and operational constraints. Values synthesised from multiple sources.}
\label{tab:cooling}
\footnotesize
\setlength{\tabcolsep}{5pt}
\begin{threeparttable}
\begin{tabularx}{\textwidth}{>{\raggedright\arraybackslash}p{2.6cm}
                             >{\centering\arraybackslash}p{1.9cm}
                             >{\centering\arraybackslash}p{1.9cm}
                             >{\centering\arraybackslash}p{2.1cm}
                             >{\raggedright\arraybackslash}X}
\toprule
\textbf{Cooling Technology} & \textbf{WUE (L/kWh)} & \textbf{Relative PUE} & \textbf{Consumptive Ratio} & \textbf{Key Constraint} \\
\midrule
Evaporative tower      & 1.99--4.00    & 1.0 (baseline) & 70--90\%       & High peak water demand; salt buildup \\[4pt]
Air-side economiser    & 0.50--1.80    & 1.05--1.15     & Variable       & Climate-dependent; humidity limits \\[4pt]
Hybrid (evap.\ assist) & 0.30--1.50    & 1.10--1.20     & 40--70\%       & Moderate water use; switchover logic \\[4pt]
Full dry cooling       & $\sim$0 direct & 1.25--1.40    & $\sim$0 direct & 25--35\% power penalty; Scope~2 water rises \\[4pt]
\bottomrule
\end{tabularx}
\begin{tablenotes}
\footnotesize
\item WUE ranges from \citet{Lei2022} and \citet{Amanambu_HSI}. PUE ratios relative to evaporative baseline
from \citet{Lei2022} and \citet{Karimi2022}. Consumptive ratios from \citet{Han2026}. Power penalties from
\citet{Han2026}. Cooling energy share from \citet{Alkrush2024}. Hybrid system WUE
from \citet{Sickinger2018}.
\end{tablenotes}
\end{threeparttable}
\end{table}

\subsection{Indirect and Embedded Water}
\label{sec:burden_indirect}

On-site cooling is only one component. Electricity-related consumption at power plants often exceeds direct facility use: the total operational footprint of U.S. data centres was $\sim$513~million m$^3$ in 2018 \citep{Siddik2021}, and off-site water intensity varies by a factor of 2.5 or more across national grids \citep{Siddik2024,Farfan2023}. Embedded supply-chain water, in semiconductor fabrication, hardware production, and construction, is the least visible component; one major operator reports that supply-chain water exceeds 99\% of its corporate footprint \citep{Apple2025}, and global AI water footprints are estimated at 312.5--764.6 billion litres for 2025 \citep{deVriesGao2026}. The hardware's material intensity reinforces this dimension: an Nvidia A100 GPU contains 32 elements including toxic metals \citep{Falk2025}. Embedded water distributes burden across geographies disconnected from the host community, from semiconductor fabrication in Taiwan to copper smelting in Chile \citep{Privette2026}. Cross-sector comparisons may mislead by conflating scopes: under 2\% of animal product water and 7--9\% of golf course water comes from public systems, whereas data centres impose thermoelectric-like demand on municipal distribution systems designed for residential loads \citep{Han2026}. The Wattnet framework shows that neglecting 15-minute electricity flow tracing misestimates both carbon and water footprints, because renewable sources that reduce carbon intensity may raise water intensity through hydropower dependency \citep{CastrillMelguizo2026}.

\subsection{The Compound Stress Problem}
\label{sec:burden_compound}

The burden pathway is most consequential when these channels coincide. Peak cooling demand temporally aligns with low streamflow, high grid stress, and peak community demand, a quadruple coincidence that is thermodynamically inevitable in hot climates. The Drought Amplification Factor formalises this, ranging from near unity on large regulated rivers to over 100 in flashy desert systems \citep{Amanambu_HSI}.

Nationally, data centre withdrawals remain at 0.6--1.1\% of total public withdrawals by 2030 \citep{Han2026}, masking severe local concentration. Independent projections converge on a rapidly growing footprint: global AI water demand could reach 4.18 million~m$^3$/day by 2030 \citep{Herrera2025}; U.S. AI server water footprints are projected at 731--1{,}125~M~m$^3$/yr for 2024--2030 \citep{Xiao2025}; and combined direct and electricity-related withdrawal could reach 4.2--6.6 billion~m$^3$ in 2027, equivalent to four to six Denmarks \citep{Li2025}. North American data centre development is projected at one trillion USD for 2025--2030 and global capex at 6.7~trillion USD through 2030 \citep{JLL2025,McKinsey2025}.

The burden pathway also operates through resource shifting: data 
centres displace other uses of finite infrastructure capacity. In 
The Dalles, Oregon, Google's water demand grew more than 300\% from 
2012 to 2024, reaching roughly one-third of the city's total supply 
and driving plans to expand reservoir capacity from Mount Hood 
National Forest \citep{OPB2026}. In Newton County, Georgia, water
costs have soared since data centre construction began, with rates
projected to increase 33\% over two years as infrastructure upgrades
are absorbed by ratepayers \citep{Shah2026}. True burden therefore 
includes the opportunity cost of foreclosed development and the 
distributional inequity of rising ratepayer costs. Individual permits 
treat each project independently, ignoring the aggregate stress 
watershed and infrastructure must absorb simultaneously.

\section{Constraint Pathways: How Water Reshapes AI Development}
\label{sec:constraint}

Table~\ref{tab:triad_synthesis} summarises the constraint pathway. Water does not merely receive burden; it determines where and how AI can grow. Treatment plants, pump stations, transmission mains, and storage cannot deliver water at the peak rates data centres require. The constraint appears through capacity bottlenecks, forced technology substitution, and regulatory access conditions, with institutional responses as downstream mediators (Table~\ref{tab:constraint_evidence}). Supplementary Table~S1 provides the complete source audit.

\subsection{Infrastructure Capacity Bottlenecks}
\label{sec:constraint_infra}

Available water in a river is not the same as deliverable water at peak demand. In Newton County, Georgia, a data centre water request of 22.7~ML/d (6~MGD) could not be accommodated because it exceeded the available surplus supply \citep{NYT2025}. In Lebanon, Indiana, full capacity of 30.3~ML/d (8~MGD) will not be available until 2031, six years after construction began, requiring a multi-year capacity expansion \citep{CityLebanon2025,Han2026}. In Port Washington, Wisconsin, a 4.5~ML/d (1.2~MGD) request triggered a 100~million USD-plus upgrade in a community with under 2~MGD of remaining capacity \citep{SpectrumNews2025}. A climate risk analysis of 1{,}455 U.S. drinking water utilities found that systems serving 67 million customers face elevated climate risk, with 36\% of high-risk utility bonds not disclosing climate exposure \citep{Lyle2025}. These are the systems now absorbing data centre connections (Supplementary Section~S2 provides extended case evidence).

\subsection{Forced Technology Substitution}
\label{sec:constraint_tech}

When capacity is insufficient, operators are pushed toward less water-intensive but less energy-efficient cooling, creating a cascading constraint. Microsoft's zero-water designs are a direct response to water constraints, not always a voluntary preference \citep{Solomon2024}. Southern Nevada has banned evaporative cooling in new developments, forcing dry cooling regardless of the 25--35\% energy penalty concentrated during peak summer periods \citep{Han2026}. This creates a paradox: water scarcity eliminates direct on-site use but \textit{increases} electricity demand, which increases electricity-related water consumption at power plants. The constraint propagates through the energy-water nexus, shifting burden from facility to grid rather than eliminating it, especially in the summer when the grid is most stressed.

\subsection{Regulatory Conditions on Infrastructure Access}
\label{sec:constraint_regulatory}

Regulatory responses are fragmented and lack standardised criteria for evaluating water burden. West Des Moines Water Works requires future projects to demonstrate peak water reduction technology \citep{Han2026}. EU Delegated Regulation 2024/1364 mandates continent-wide WUE disclosure \citep{EU2024}. The Susquehanna River Basin Commission requires environmental review for data centre withdrawals \citep{SRBC2025}. California and Texas mandate water use reporting. Singapore's temperature mandates indirectly constrain cooling choice \citep{SPGlobal2025}. Thames Water has discussed restricting peak-period use \citep{Gorey2025}. Collectively, this patchwork represents water governance constraining AI deployment.

WUE reporting is mandated by several jurisdictions \citep{EU2024,SingaporeGreenDC2024,GermanyEnEfG2023}, but does not sufficiently reflect water use for water resources planning or sustainability. Its IT energy denominator creates a perverse incentive: operators increasing computational intensity can worsen absolute water consumption while improving their WUE score \citep{Privette2026}. WUE is location-agnostic, failing to distinguish freshwater from recycled sources, and its seasonal variability (1.3--2.5~L/kWh within one facility) permits selective reporting \citep{Privette2026}. Singapore mandates WUE targets of 2.0~L/kWh for new data centres under its Green Data Centre Roadmap \citep{SingaporeGreenDC2024}, while Germany's Energy Efficiency Act requires WUE reporting but does not yet set a binding limit \citep{GermanyEnEfG2023}. Without complementary indices for water type and geographic context, however, such measures may misdirect rather than constrain actual impact.

Information asymmetry is a parallel constraint. Operators and host jurisdictions
have resisted disclosure of facility-level data, claiming trade-secret status
\citep{VirginiaMercury2024}. Citizen groups including the Dorchester County FOIA
case against Google \citep{PostCourier2024} and the Milwaukee Riverkeeper lawsuit
against Microsoft in Racine County \citep{WPR_Racine2025} have been required to
compel basic disclosure, undermining democratic water governance \citep{Shah2026}.

Physical constraints quickly become institutional and financial. Developer-funded infrastructure investments now range from 25~million to over 400~million USD per project \citep{VirginaDEQ2024,Amazon2026,CityLebanon2025}. Where data centres depend on transboundary supply, as in Singapore's reliance on Malaysia \citep{SPGlobal2025}, physical constraints are coupled with supply-chain vulnerabilities.

\begin{table}[H]
\centering
\caption{Evidence of infrastructure constraining AI development.}
\label{tab:constraint_evidence}
\scriptsize
\setlength{\tabcolsep}{4pt}
\renewcommand{\arraystretch}{1.08}
\begin{threeparttable}
\begin{tabularx}{\textwidth}{>{\raggedright\arraybackslash}p{2.2cm}
                              @{\hspace{6pt}}>{\raggedright\arraybackslash}p{2.2cm}
                              >{\raggedright\arraybackslash}X
                              >{\raggedright\arraybackslash}p{3.0cm}}
\toprule
\textbf{Mechanism} & \textbf{Location} & \textbf{Description} & \textbf{Consequence \& Source} \\
\midrule
\multicolumn{4}{l}{\textit{Infrastructure constraint mechanisms}} \\[1pt]
Capacity bottleneck & Newton County, GA   & Water request exceeds available surplus                       & Project blocked \citep{NYT2025} \\
Capacity bottleneck & Lebanon, IN         & 30.3~ML/d (8~MGD) not available until 2031; multi-year capacity expansion & 6+ year delay \citep{CityLebanon2025} \\
Capacity bottleneck & Port Washington, WI & 4.5~ML/d (1.2~MGD) request; $<$7.6~ML/d ($<$2~MGD) remaining capacity                 & \$100M+ upgrade \citep{SpectrumNews2025} \\
Technology forcing  & Southern Nevada     & Evaporative cooling banned in new developments                & 25--35\% power penalty \citep{SPGlobal2025} \\
Technology forcing  & Microsoft (global)  & Zero-water DC designs adopted as constraint response          & Scope~2 water shifts to grid \citep{Solomon2024} \\
Regulatory condition & West Des Moines, IA & Utility requires peak reduction technology on future projects & Future projects conditioned \citep{Han2026} \\
Regulatory condition & EU-wide            & Delegated Reg.\ 2024/1364 mandates WUE disclosure             & WUE disclosure required \citep{EU2024} \\
Regulatory condition & Singapore          & Temperature mandates indirectly constrain cooling choice      & Indirect technology forcing \citep{SPGlobal2025} \\
\midrule
\multicolumn{4}{l}{\textit{Institutional mediators (downstream)}} \\[1pt]
Economic      & Leesburg, VA & \$25M water infrastructure upgrade                    & Developer-funded \citep{VirginaDEQ2024} \\
Economic      & Louisiana    & \$400M water infrastructure investment                 & Corporate investment \citep{Amazon2026} \\
Transboundary & Singapore    & Dependence on Malaysian supply; agreement expires 2061 & Supply-chain risk \citep{SPGlobal2025} \\
\bottomrule
\end{tabularx}
\end{threeparttable}
\end{table}

\section{Adaptive Pathways: AI-Enabled Water System Adaptation}
\label{sec:adaptive}

Table~\ref{tab:triad_synthesis} summarises the adaptive pathway, and Supplementary Table~S2 catalogues the real-world deployment evidence on which this section draws. This section asks a narrower question than most AI-for-water literature: not whether AI can improve water systems in general, but whether those improvements can offset the specific burdens imposed by co-located data centres in the same places, at the same times, and at usable scales. Evidence that AI \textit{can} improve water systems is strong and growing; evidence that AI-driven improvements \textit{are already offsetting} co-located data centre burdens is essentially absent.

\subsection{Leak Detection and Non-Revenue Water Recovery}
\label{sec:adaptive_leaks}

Leak reduction is the clearest adaptive pathway because it directly recovers usable capacity within existing systems. The United States loses about 19.5\% of treated drinking water through distribution leaks and other non-revenue sources \citep{BluefieldNRW2025}, an estimated $\sim$6~Bgal/day of non-revenue water nationally, substantially exceeding projected 2030 data centre withdrawal \citep{Han2026}.

AI-driven leak detection is now mature. Multi-gene genetic programming has demonstrated operationally useful leakage prediction \citep{Hayslep2025}, and deep learning models have shown transferability across multiple distribution systems \citep{Jun2025}. A systematic review of 53 studies (2018--2025) confirms convolutional neural networks achieve 95--99\% accuracy and support vector machines 94--100\% \citep{ZunigaUribe2026}. A bibliometric analysis of 600 articles (2000--2023) traces the evolution from manual sounding to sensor-driven and AI-enabled approaches, with pressure-based anomaly detection the most sensitive input \citep{Farah2024}. A state-of-the-art review spanning scientific, industrial, and commercial developments estimates global water loss at 126 billion m$^3$/yr and documents the convergence of machine learning, patented hardware, and commercial platforms into increasingly integrated and scalable monitoring strategies \citep{Sousa2026}. A comprehensive machine learning review documents capabilities spanning demand forecasting, leak detection, and pipe condition assessment \citep{Taloma2025}. An AI-enhanced digital twin reports 94.2\% detection accuracy and 15\% loss reduction in pilots \citep{WaterTwinAI2025}. Graph neural networks are promising for distribution network modelling, spanning leak detection, demand forecasting, sensor placement, and quality monitoring \citep{Vittori2025}, with physics-informed variants enabling scalable pressure estimation without extensive labelled data \citep{GNN_WDS2025}.

Real deployments provide preliminary evidence of scale. Thames Water deployed permanent acoustic sensors across its distribution network, achieving measurable leakage reductions in pilot zones and contributing to the utility's 15\% leakage reduction target \citep{WaterBriefing2023}. Comprehensive evidence of sustained, system-wide savings remains limited.

\textit{Offset conditions.} Leak recovery can accommodate data centre demand without new source development only if it occurs within the same distribution system, during the same peak periods, and the recovered volume is usable. In small systems with ageing infrastructure, which are most stressed by data centre demand, leak rates are highest but implementation capacity is lowest. No study has tested whether recovered water can offset co-located data centre demand during peak periods.

\subsection{Demand Forecasting for Peak Management}
\label{sec:adaptive_demand}

Demand forecasting matters because the binding constraint is peak capacity, not annual average. Deep learning in urban water management supports short-term (hourly to daily) and medium-term (weekly to monthly) forecasting \citep{Fu2022}. Real-time demand estimation enables dynamic allocation during peak periods when cooling, residential irrigation, and fire-flow reserves compete for the same capacity. Coupled with water-aware scheduling, forecasting creates closed-loop optimisation: the utility predicts tomorrow's peak, signals the data centre, and the data centre shifts deferrable workloads. The WATCH algorithm showed that water-constrained geographic load balancing can reduce total consumption by $\sim$20\% with negligible cost, by shifting workloads from water-stressed to water-abundant locations in real time \citep{Islam2015}. Google's Carbon-Intelligent Compute Management provides an operational template directly transferable from carbon to water constraints \citep{Radovanovic2022}.

\textit{Offset conditions.} Demand forecasting does not create new water; it enables efficient allocation of existing capacity. The offset is operational (reduced peak shortage) rather than volumetric.

\subsection{Digital Twins for Infrastructure Stress Testing}
\label{sec:adaptive_twins}

Digital twins bring the river and pipe questions into one planning environment: what happens when a 500~MW data centre connects to a system with 7.6~ML/d (2~MGD) surplus, or when heatwave, drought, and peak demand coincide? An AI-enhanced digital twin combining LSTM, Prophet, and XGBoost ensembles achieves 94.2\% prediction accuracy \citep{WaterTwinAI2025}. A review of 147 water-sector digital twin studies finds only 8--9 full-scale deployments to date, with documented gains including 30\% aeration energy reduction and 16.7\% pumping electricity savings, though most remain advisory rather than autonomous \citep{GhorbaniBam2025}. For data centre siting, digital twins could integrate HSI and infrastructure capacity analysis, providing a unified ``Can the river handle it?''/``Can the pipes handle it?'' tool that currently exists in neither discipline \citep{Amanambu_HSI,Han2026}.

\textit{Offset conditions.} Digital twins do not save water; they enable better decisions. Their value is preventing compound stress events rather than recovering lost water.

\subsection{Streamflow Forecasting and Drought Early Warning}
\label{sec:adaptive_streamflow}

Streamflow forecasting matters because adaptive response is only useful if drought can be anticipated early enough to change operations \citep{Kratzert2019,Nearing2021}. The National Water Model provides continental-scale flow forecasts that could underpin water-aware scheduling. Within the HSI framework, day-to-week forecasts could enable curtailment triggers: when predicted drought flow falls below thresholds pushing HSI above critical levels, data centres could shift workloads to hydrologically safer locations \citep{Amanambu_HSI}, directly coupling adaptive and constraint pathways. Google's flood forecasting now delivers real-time warnings in over 80 countries, demonstrating that continental-scale streamflow intelligence is technically feasible; adapting it for drought curtailment is institutionally rather than technically constrained.

\textit{Offset conditions.} Forecasting enables preemptive response only if workloads are deferrable or relocatable, lead times suffice, and institutional mechanisms translate forecasts into curtailment signals.

\subsection{Treatment Optimisation and Reclaimed Water}
\label{sec:adaptive_treatment}

Treatment optimisation and reclaimed water substitute new usable capacity for new withdrawals. Deep reinforcement learning has been applied to chemical dosing, energy management, and process control \citep{Negm2024}. Reclaimed water is a promising nexus between burden and adaptive pathways: several major data centres use reclaimed wastewater for cooling, and AI can optimise reclamation by monitoring quality and adjusting treatment parameters dynamically.

Operator deployments are moving beyond pilots. AWS uses recycled water at 24 data centre locations globally, including facilities in Oregon, Virginia, and Singapore, and reports recycling over $\sim$2{,}000~ML ($\sim$530~million gallons) across its operations \citep{Amazon2025recycled}. This demonstrates that infrastructure and regulatory barriers to non-potable reuse are surmountable at scale when operators invest in treatment upgrades and negotiate agreements with municipal authorities. The AWS cases illustrate the importance of early engagement: at several sites, reuse agreements were negotiated during siting rather than retrofitted, reducing both cost and permitting delays \citep{Amazon2025recycled}.

\textit{Offset conditions.} Reclaimed water substitution reduces potable demand one-for-one but requires proximity to a wastewater plant with surplus capacity, acceptable water chemistry, and regulatory approval. The U.S. EPA's Water Reuse Action Plan 2.0, launched in April 2026, explicitly targets data centre cooling and microchip fabrication as priority sectors for accelerated water reuse \citep{EPAWRAP2026}, signalling federal policy alignment with this adaptive pathway.

\subsection{Co-Location Synergies and Waste Heat Recovery}
\label{sec:adaptive_colocation}

Co-location synergies represent the most complete but most site-specific version of the adaptive pathway. Data centre waste heat can warm influent to wastewater treatment plants, improving biological treatment efficiency in cold climates \citep{Chen2018}. A campus-scale EPFL analysis shows data centres modelled as heat-active prosumers can supply up to 40\% of district heating via Organic Rankine Cycle waste heat recovery, converting a thermal burden into a community energy resource \citep{Ravi2026}. Learning-augmented online control for decarbonising water infrastructure shows AI can optimise the energy balance of water systems accounting for co-located heat sources \citep{Yang2024}. These co-designs represent the most complete closure of the loop: burden (stress), constraint (forced co-location), and adaptation (AI optimisation) operating together.

\textit{Offset conditions.} Co-location synergies require proximity, compatible thermal demands, and institutional arrangements. They are site-specific rather than generalisable.

\subsection{The Conditional Nature of Adaptation}
\label{sec:adaptive_synthesis}

The evidence reviewed in Sections~\ref{sec:adaptive_leaks}--\ref{sec:adaptive_colocation} supports a conditional, rather than automatic, offset story (Table~\ref{tab:offset_conditions}). Whether AI-driven efficiency can offset data centre demand at a locally significant scale remains an untested hypothesis; it would require intentional design and site-specific conditions to be met. The same industry stressing water systems is building the tools to improve them; closing the loop requires moving from aggregate potential to facility- and community-level matching of burden and benefit.

A survey of 64 water utilities across 28 countries measures the implementation gap: the median Utility Digitalisation Score (UDS), which rates adoption of digital technologies from 0 (not planned) to 3 (in operation), for distribution networks is only 2.20 out of 3, with wastewater operations and customer-demand management each at 1.83 \citep{Daniel2023}. The gap between demonstrated digital and AI capability and actual deployment is widest in the small and medium utilities hosting most new data centre connections. No peer-reviewed study has demonstrated that AI-driven water savings have offset co-located data centre burden at operational scale.

\begin{table}[H]
\centering
\caption{Adaptive pathway mechanisms and the conditions required for meaningful offset of data center water burden.}
\label{tab:offset_conditions}
\footnotesize
\setlength{\tabcolsep}{4pt}
\renewcommand{\arraystretch}{1.1}
\begin{threeparttable}
\begin{tabularx}{\textwidth}{>{\raggedright\arraybackslash}p{2.0cm}
                              >{\raggedright\arraybackslash}p{2.0cm}
                              >{\raggedright\arraybackslash}X
                              >{\raggedright\arraybackslash}p{3.8cm}}
\toprule
\textbf{Mechanism} & \textbf{Offset type} & \textbf{Key conditions for local matching} & \textbf{Maturity} \\
\midrule
Leak detection
  & Volumetric (recovered supply)
  & Same distribution system; recovered during peak; implementation capacity in small systems
  & Detection accuracy validated across 53 studies (95--99\%); system-wide NRW recovery demonstrated in pilots but not at scale \\[2pt]
Demand forecasting
  & Operational (better allocation)
  & Real-time data integration; institutional mechanisms for signals
  & Technically mature; institutionally nascent \\[2pt]
Digital twins
  & Decision support (avoided stress)
  & Sufficient data for calibration; planning integration
  & 8--9 full-scale deployments documented; most advisory rather than autonomous \\[2pt]
Streamflow forecasting
  & Preemptive scheduling
  & Deferrable workloads; sufficient lead time; curtailment governance
  & Technically mature; governance absent \\[2pt]
Reclaimed water
  & Volumetric (substitution)
  & Proximity to treatment plant; quality compatibility; regulatory approval
  & Operational at multiple sites \\[2pt]
Co-location synergies
  & System efficiency (thermal)
  & Physical proximity; compatible thermal demands; institutional arrangements
  & Concept stage; few operational examples \\
\bottomrule
\end{tabularx}
\end{threeparttable}
\end{table}

\section{Integrated Synthesis: Feedback Regimes, Water Consumption Impact, and Host Community Typologies}
\label{sec:synthesis}

The three pathways converge on three analytical constructs: an account of how they interact, a quantitative framework for assessing net local effect, and a typology for comparing host communities. Table~\ref{tab:quant_synthesis} collects the headline quantitative indicators for each pathway, and the integrated feedback architecture is previewed in Figure~\ref{fig:loops}.

\subsection{Divergent Feedback Regimes}
\label{sec:synthesis_regimes}

The three pathways interact as feedback regimes whose trajectory depends on whether coupling is managed. Two bounding trajectories emerge. Without coordination, each pathway amplifies the others: AI growth raises withdrawals and peak demand; water stress triggers bottlenecks; constraints force dry cooling despite the 25--35\% power penalty; higher power raises indirect water use; grid strain and scarcity push siting into progressively less suitable locations; worse locations amplify stress (location dominates by $\sim$4{,}897$\times$). The cycle returns with a higher baseline stress. This regime is already observable: constrained Virginia sites are expanding into the Great Lakes, competing with critical minerals mining and irrigation for the same finite water resources \citep{AllianceGreatLakes2025}.

With intentional coupling, AI tools improve water-system efficiency, freed capacity enables efficient evaporative cooling, power demand falls by 25--35\%, indirect water use declines, and continued investment reduces baseline stress. The cycle returns with lower net burden. However, reinforcing burden is the default, requiring no coordination, while enabling adaptation requires shared data platforms, water-aware workload scheduling, and regulatory frameworks that treat data centres as a distinct user class. Barriers are primarily technical and institutional: data silos between utility SCADA and data centre cooling controllers; mismatched planning horizons (20--50 years for utilities versus 18--36 months for data centre build cycles); and the absence of standardised community-scale burden indicators.

\subsection{Water Consumption Impact: A Community-Scale Index for Utility Impact}
\label{sec:synthesis_net}

Impacts that appear modest nationally can produce acute burden at the community scale, because aggregate indicators, including Water Positive pledges and facility-level WUE, average across time, space, and institutional boundaries. National analyses such as \citet{Han2026} quantify the macro problem and identify affected communities, but do not provide a standardised framework for comparing burden severity across sites. A useful assessment must therefore operate at the site level, expressing each facility's cooling water demand as a fraction of the host utility's capacity to deliver it.

We focus on \textit{community-scale burden rather than watershed-scale scarcity}. The index captures how strongly a facility's peak consumptive demand presses against the host utility's service capacity, isolating the dimension that determines whether a community can physically serve the facility on its worst day. We introduce the \textit{Water Consumption Impact} (WCI) index, the fraction of host utility peak-day capacity permanently consumed by a facility's evaporative cooling demand:
\begin{equation}
\mathrm{WCI} = \frac{C_{\mathrm{peak}}}{K}
\label{eq:wci}
\end{equation}
where $C_{\mathrm{peak}}$ is the facility's peak-day consumptive water use (ML/d) and $K$ is the host public water system's maximum deliverable capacity (ML/d). WCI is bounded below by zero and unbounded above; values exceeding unity indicate that a single facility's peak cooling demand exceeds the total capacity of its host utility.

A three-factor decomposition makes the drivers transparent:
\begin{equation}
\mathrm{WCI} = \frac{W}{K} \times r \times \mathrm{PF}
\label{eq:wci_decomp}
\end{equation}
Three quantities follow immediately. First, $W$ is the facility's average daily withdrawal (ML/d) and $W/K$ is the \textit{demand scale}: the facility size relative to the host utility, a siting decision and the dominant driver of cross-site variation. Second, $r$ is the \textit{consumptive ratio}: the fraction of withdrawn water lost to evaporation, a cooling-technology parameter (typically 0.70--0.90 for evaporative systems). Third, $\mathrm{PF}$ is the \textit{peaking factor}: peak-day to average-day demand, from 1.0 for steady-state facilities to over 10 for those with extreme seasonal swings. Both $r$ and $\mathrm{PF}$ are dimensionless, so $W$, $C_{\mathrm{avg}}$, and $C_{\mathrm{peak}}$ share the same units (ML/d). The intermediates $C_{\mathrm{avg}} = r \times W$ and $C_{\mathrm{peak}} = C_{\mathrm{avg}} \times \mathrm{PF}$ are defined in Supplementary Eqs.~(S1)--(S2).

Each factor maps to a distinct policy lever, making WCI diagnostic and actionable. Demand scale is set by the siting decision and is most difficult to change after construction. The consumptive ratio is governed by cooling technology, with dry cooling eliminating consumptive use at the cost of the 25--35\% power penalty \citep{Han2026}. The peaking factor reflects operational demand patterns and is most amenable to demand-side management through workload scheduling and thermal storage. WCI therefore tells communities and regulators not only \textit{how severe} the burden is but \textit{which lever to pull}.

WCI embeds and extends WUE. Because $W = \mathrm{WUE} \times E_{\mathrm{IT}}$ under the standard Green Grid definition (WUE $=$ annual site water usage / IT equipment energy, L/kWh), WCI can be written as $\mathrm{WCI} = (\mathrm{WUE} \times E_{\mathrm{IT}} / K) \times r \times \mathrm{PF}$. WUE measures facility efficiency; WCI adds the host-system denominator $K$ that anchors the index to community-level utility burden. A low-WUE facility can still produce high WCI in a capacity-constrained community, and vice versa . WUE's IT-energy denominator creates perverse incentives, its location-agnostic formulation obscures geographic stress, and its seasonal variability permits selective reporting. WCI addresses each limitation.

To translate consumption into community terms, we define a household equivalence measure:
\begin{equation}
\mathrm{HH}_{\mathrm{equiv}} = \frac{C_{\mathrm{avg}} \times 365.25}{H_{\mathrm{avg}}}
\label{eq:hh_equiv}
\end{equation}
where $H_{\mathrm{avg}} = 0.4146$~ML/household/year $\sim$1{,}136~L/day/household ($\sim$300~gallons/day); \citep{EPA_WaterSense2024}). Per capita domestic water use is 310~L/person/day (82~gallons/person/day) \citep{USGS2018}. This converts a volumetric flow into a tangible community burden.

Because AI is growing rapidly, a static snapshot understates the trajectory of local stress. We project WCI forward:
\begin{equation}
\mathrm{WCI}(t) = \mathrm{WCI}_{0} \times (1 + g)^{t}
\label{eq:wci_growth}
\end{equation}
with $g = 0.13$, the lower bound of projected U.S. data centre energy growth (13--27\% annually, reaching 325--580~TWh by 2028; \citep{Shehabi2024}). This rate is deliberately conservative: EPRI's February 2026 update, based on commercial development data rather than chip-shipment proxies, projects 380--790~TWh by 2030, roughly 60\% above its own 2024 estimates \citep{EPRI2026}, and Microsoft's reported water consumption grew at approximately 24\% CAGR (2021--2023). Energy growth serves as a water proxy because cooling water scales with IT heat load. For illustration, at $g = 0.20$ (closer to the EPRI Medium scenario), Botetourt County would breach 100\% by 2030 rather than 2033, and Mayes County by 2027 rather than 2029, shortening planning horizons by two to three years. The projection identifies the year each site breaches the $\mathrm{WCI} = 100\%$ threshold, providing a planning horizon.

Equations~\ref{eq:wci}--\ref{eq:wci_growth} form the core framework. Supporting equations are in Supplementary Eqs.~(S1)--(S3); input data are in Table~\ref{tab:wci_input}; host-utility capacities and computed results are in Supplementary Tables~S5--S6.

\subsubsection{Applying the Water Consumption Impact index to ten US data centres}
\label{sec:wci_results}

We apply WCI to ten US data centre locations where utility capacity and facility demand are assembleable from public sources. Inputs come from operator environmental reports \citep{Google2025ER}, public utility records, permit filings, and \citet{Han2026}. Site-specific withdrawal ($W_d$), consumption ($C$), and PUE for the six Google campuses (Council Bluffs, Mayes Co., The Dalles, Douglas Co., Midlothian, Henderson) are taken from the Google 2025 Environmental Report (p.~110, FY2024 data independently assured by EY), with $r = C/W$ computed site by site rather than assumed at fleet average. For non-Google operators, $r = 0.77$ from \citet{Han2026}. Table~\ref{tab:wci_input} summarises inputs for Figures~\ref{fig:wci_panel_a}--\ref{fig:wci_growth}; $K$ and context are in Supplementary Table~S5, computed results in Supplementary Table~S6.

\begin{table}[H]
\centering
\caption{Input data for the WCI analysis across ten US data centre locations. $W_d$ is average daily withdrawal, $r$ is consumptive ratio, PF is peaking factor, PUE is Power Usage Effectiveness (where available from operator reports). $C_{\mathrm{avg}} = r \times W_d$; $C_{\mathrm{peak}} = C_{\mathrm{avg}} \times \mathrm{PF}$. Symbol legend: ``---'' $=$ absent from accessible public disclosure; $\diamond$ $=$ default value applied in the absence of a site-specific disclosure; $\blacktriangle$ $=$ value derived analytically in an upstream source rather than measured directly at the site.}
\label{tab:wci_input}
\scriptsize
\setlength{\tabcolsep}{3pt}
\begin{threeparttable}
\begin{tabular}{@{}l l c c c c c c >{\raggedright\arraybackslash}p{2.6cm}@{}}
\toprule
\textbf{Location} & \textbf{Operator} & \textbf{$W_d$} & \textbf{$r$} & \textbf{PF} & \textbf{PUE} & \textbf{$C_{\mathrm{avg}}$} & \textbf{$C_{\mathrm{peak}}$} & \textbf{Source} \\
 & & (ML/d) & & & & (ML/d) & (ML/d) & \\
\midrule
Lebanon, IN          & Meta      & 4.81$\blacktriangle$  & 0.77\tnote{b}  & 6.3$\blacktriangle$\tnote{c}  & ---            & 3.70  & 23.32 & \citet{Han2026}, App.~D; \citet{CityLebanon2025}\tnote{d} \\
Council Bluffs, IA   & Google    & 14.62 & 0.716\tnote{a} & 6.5$\blacktriangle$\tnote{c}  & 1.09\tnote{a}  & 10.47 & 68.02 & \citet{Google2025ER}, p.~110\tnote{a} \\
Mayes Co., OK        & Google    & 11.49 & 0.752\tnote{a} & 2.5$\diamond$\tnote{e}  & 1.11\tnote{a}  & 8.64  & 21.59 & \citet{Google2025ER}, p.~110\tnote{a} \\
The Dalles, OR       & Google    & 4.78  & 0.784\tnote{a} & 2.21\tnote{c} & 1.08\tnote{a}  & 3.75  & 8.28  & \citet{Google2025ER}, p.~110\tnote{a} \\
Douglas Co., GA      & Google    & 4.60  & 0.826\tnote{a} & 2.5$\diamond$\tnote{e}  & 1.09\tnote{a}  & 3.80  & 9.50  & \citet{Google2025ER}, p.~110\tnote{a} \\
Wisconsin site       & Microsoft & 0.087 & 0.77\tnote{b}  & 30\tnote{f}   & ---            & 0.067 & 2.01  & \citet{WisPubRadio2025}; \citet{Han2026}, \S3.1.2\tnote{f} \\
Botetourt Co., VA    & Google    & 7.57  & 0.77\tnote{b}  & 2.5$\diamond$\tnote{e}  & ---            & 5.83  & 14.57 & \citet{GoogleBotetourtFAQ}\tnote{g} \\
Memphis, TN          & xAI       & 3.79  & 0.77\tnote{b}  & 4.5$\diamond$\tnote{h}  & ---            & 2.91  & 13.12 & \citet{GEAA2025}\tnote{i} \\
Midlothian, TX       & Google    & 2.29  & 0.825\tnote{a} & 2.5$\diamond$\tnote{e}  & ---            & 1.89  & 4.72  & \citet{Google2025ER}, p.~110\tnote{a} \\
Henderson, NV        & Google    & 3.73  & 0.576\tnote{a} & 2.5$\diamond$\tnote{e}  & 1.09\tnote{a}  & 2.15  & 5.37  & \citet{Google2025ER}, p.~110\tnote{a} \\
\bottomrule
\end{tabular}
\begin{tablenotes}
\scriptsize
\item[a] \citet{Google2025ER}: site-specific withdrawal, consumption, and PUE reported for each Google campus on p.~110 (FY2024 data, independently assured by EY); $r$ computed directly as $r = C/W$.
\item[b] For non-Google operators (Meta, Microsoft, xAI) and for Botetourt County, $r = 0.77$ is assigned as the weighted mean reported in \citet{Han2026}, App.~A.5, consistent with the 0.70--0.90 range for evaporative cooling.
\item[c] Peaking factors: Lebanon PF $=6.3$ from \citet{Han2026} App.~D (analytical recomputation, $\blacktriangle$). Council Bluffs PF $=6.5$ from \citet{Han2026} \S3.1.2, applying a 1.5 regulatory safety factor to the measured 4.30 monthly peak ratio. ($\blacktriangle$) and the Dalles PF $=2.21$.
\item[d] Lebanon current-phase intake 4.81~ML/d (1.27~MGD) corresponds to \citet{Han2026} App.~D recomputation ($\blacktriangle$); phased buildout to 30.3~ML/d (8~MGD) by 2031 per \citet{CityLebanon2025}.
\item[e] Default PF $= 2.5$ applied to warm-climate sites where no site-specific peaking factor has been reported ($\diamond$).
\item[f] The Wisconsin hyperscale site is a Microsoft data centre in Mount Pleasant / Racine County, WI. Operating figures (${\sim}31.8$~ML/year (${\sim}8.4$~million gallons/year); peak demand ${>}30\times$ the average-day equivalent) are from \citet{WisPubRadio2025}; the PF${\approx}30$ figure is also adopted in \citet{Han2026} \S3.1.2.
\item[g] \citet{GoogleBotetourtFAQ}: 2~MGD permitted reservation via Western Virginia Water Authority; facility still ramping, $r$ assumed.
\item[h] xAI Memphis (Colossus) PF $=4.5$ is a placeholder reflecting the thermal load profile of a large AI training cluster ($\diamond$).
\item[i] \citet{GEAA2025}: xAI Memphis consumes approximately 3.79~ML/d (1~MGD) at current operations.
\end{tablenotes}
\end{threeparttable}
\end{table}

\paragraph{How severe is the burden, and where does it fall?} Across the ten sites, WCI spans nearly three orders of magnitude (Figure~\ref{fig:wci_panel_a}a).
At the low end, Henderson, Nevada registers WCI of 0.002 because the regional Southern Nevada system dwarfs any single facility. At the high end, Lebanon, Indiana registers 1.34, meaning peak-day consumptive demand already exceeds the host utility's full 17.4~ML/d capacity. Even measured against the planned 25~MGD (94.6~ML/d) dedicated supply for the development zone \citep{CityLebanon2025}, WCI remains 0.25, meaning a single facility's peak cooling demand would claim one quarter of the new purpose-built capacity. The community-share (Figure~\ref{fig:wci_panel_a}b) shows that at several sites, cooling demand approaches the residential water use of the surrounding community, even though the national aggregate contribution remains modest.

\begin{figure}[H]
\centering
\includegraphics[width=\textwidth]{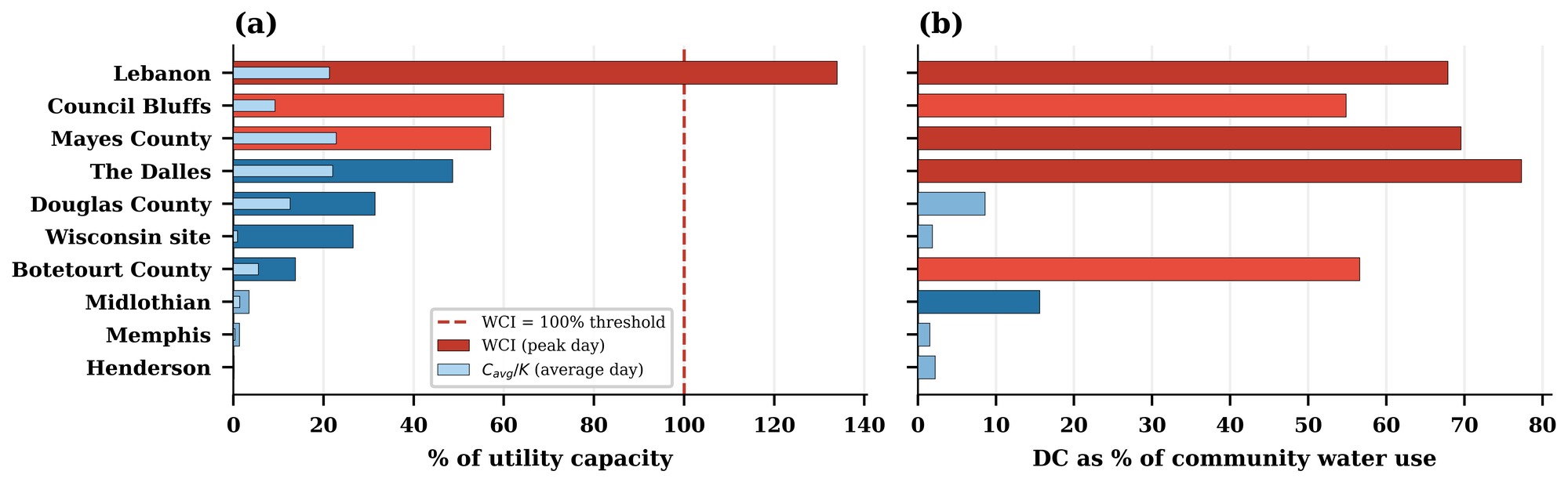}
\caption{\textbf{Water Consumption Impact and community water share across ten US data-centre locations.} (a)~Peak-day WCI for each site, expressed as a percentage of host-utility delivery capacity. The red dashed line marks the 100\% threshold; Lebanon, IN crosses it, meaning that on the worst day its cooling demand alone would claim the entire deliverable capacity of the host system. The lighter bars show the same sites on an average-day basis ($C_{\mathrm{avg}}/K$), separating routine load from peak-day amplification. (b)~Facility water consumption expressed as a share of total community water use. Several small hosts show cooling demand that rivals the residential water consumption of their surrounding communities. }
\label{fig:wci_panel_a}
\end{figure}

\paragraph{Which factor drives each site's burden?} Decomposing WCI into its three components via Eq.~\ref{eq:wci_decomp} (Figure~\ref{fig:wci_panel_b}a) exposes the site-specific geometry. Lebanon, Council Bluffs, and the Wisconsin (Microsoft) site are dominated by the peaking factor \citep{Han2026,WisPubRadio2025}. Mayes County and The Dalles are dominated by demand scale ($W/K$), and Douglas County by its consumptive ratio because its warm-humid climate pushes the Google-reported value to $r = 0.826$ \citep{Google2025ER}. No single factor binds universally, so the right intervention depends on what is driving the local number. Community equivalence (Eq.~\ref{eq:hh_equiv}) places these ratios on a human scale (Figure~\ref{fig:wci_panel_b}b): Lebanon consumes roughly 3{,}261 household-equivalents, The Dalles $\sim$3{,}301, Mayes County $\sim$7{,}610, and Council Bluffs the largest absolute footprint at $\sim$9{,}220, even though its WCI is moderate. Fractional burden and absolute volume tell different halves of the same story.

\paragraph{Which lever should each site pull?} Each factor maps to a distinct policy lever. \textit{Siting} acts on $W/K$ via smaller facilities, larger hosts, or phased buildout. \textit{Technology} acts on $r$ via lower-evaporative cooling, bounded by the 25--35\% energy penalty of dry cooling \citep{Han2026}. \textit{Management} acts on PF via workload scheduling, thermal storage, and demand-side contracts. We recompute WCI with each lever applied in isolation: siting caps $W/K \to 0.75 \times W/K$, technology caps $r \to \min(r, 0.25)$, and management caps $\mathrm{PF} \to \min(\mathrm{PF}, 2.0)$ (Figure~\ref{fig:wci_growth}a). The siting lever returns a flat 25\% everywhere because it scales with $W$. Technology dominates at sites with already-high $r$ but moderate peaking (Douglas County, Mayes County, The Dalles, Botetourt County, Midlothian, Memphis, Henderson). Management dominates where PF is the binding constraint (Lebanon, Council Bluffs, Wisconsin). Technology-dominated sites cannot be solved by demand management alone, and vice versa; the framework points each community to its own binding lever rather than prescribing a single policy.

\paragraph{How much time do these communities have?} Figure~\ref{fig:wci_growth}b projects WCI to 2035 under $g = 0.13$ \citep{Shehabi2024} with $K$ constant. Planning horizons are short: Council Bluffs crosses 100\% by 2029, Mayes County by 2029, The Dalles by 2030, Douglas County by 2034, and the Wisconsin site by 2035, while Lebanon continues to diverge from an already-unsustainable baseline. Two low-burden sites (Memphis, Henderson) are omitted from the panel for visual contrast. Projections are conservative: they hold $K$ fixed, assume no additional co-location, and exclude climate-driven temperature rise. Relaxing any of these would move breach dates forward. For context, total U.S. electricity demand is forecast to grow by only 1.2\% in 2026 and 3.3\% in 2027 \citep{EIA_STEO2026}, yet the commercial sector, which includes data centres leads this growth, with summer demand rising 6\% by 2027, underscoring the disproportionate role of data centres in shaping regional infrastructure stress.

\begin{figure}[H]
\centering
\includegraphics[width=\textwidth]{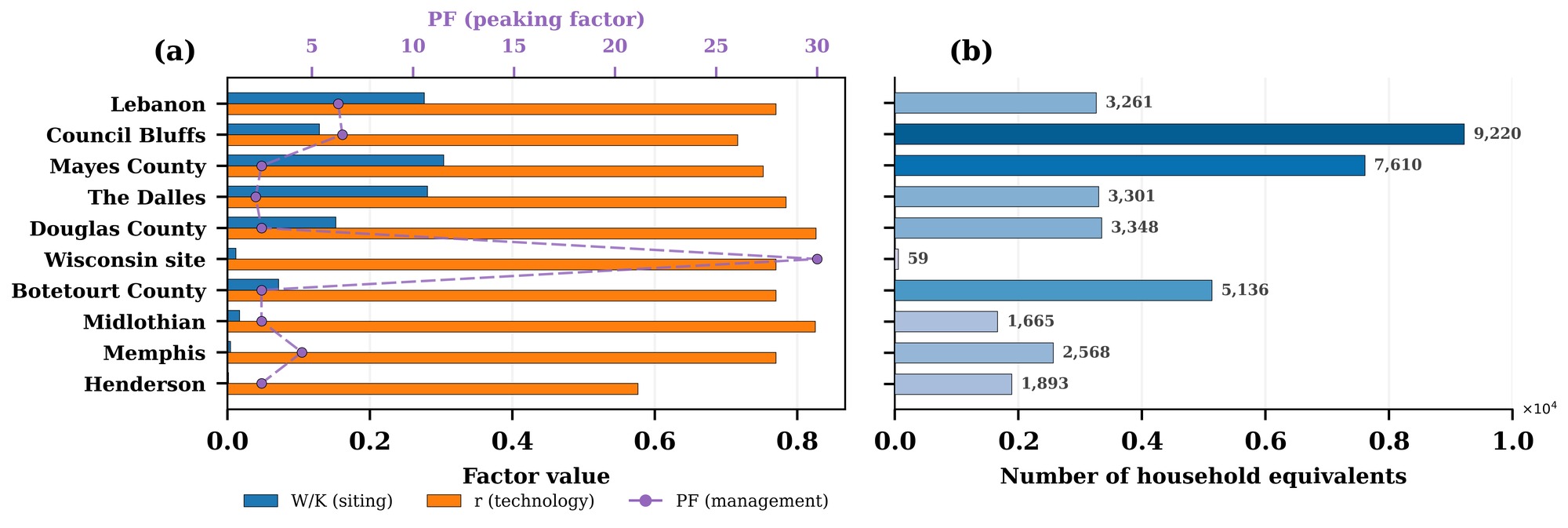}
\caption{\textbf{Three-factor decomposition of WCI and household-equivalent footprints across ten US data-centre locations.} (a)~WCI is decomposed into its three drivers: demand scale $W/K$ (blue bars) and consumptive ratio $r$ (orange bars) on the bottom axis, and peaking factor PF (purple line with circle markers) on the top axis. The bars show why the binding factor is site-specific: PF dominates at Lebanon, Council Bluffs, and the Wisconsin site; $W/K$ dominates at Mayes County and The Dalles; $r$ dominates at Douglas County. (b)~Household equivalence: the number of U.S. households whose annual water consumption matches one facility's average-day cooling demand, with darker bars indicating larger absolute footprints. }
\label{fig:wci_panel_b}
\end{figure}

\FloatBarrier
WCI complements, rather than competes with, the national analysis of \citet{Han2026}, who project 2,638--5,493~ML/d (697--1{,}451~MGD) of new capacity and \$7--58~B in infrastructure costs by 2030. The two operate at different scales: Han et~al.\ answer ``how much, nationally?'' from aggregated industry data, while WCI answers ``how severe is the burden in this community, which factor drives it, and when will it breach capacity?'' at the individual host. National projections motivate urgency; community-scale assessment identifies where intervention is needed and which lever will move the number.

\paragraph{Scale contrast: applying the framework to Northern Virginia.} To illustrate the importance of community scale, consider the Potomac River basin---the world's largest concentration of data centres, with $\sim$5{,}400~MW of aggregate IT capacity across more than 290 facilities \citep{ICPRB2026}. In 2026, American Rivers designated the Potomac as the most endangered river in the United States, citing the rapid, unchecked buildout of data centres as a primary threat to the drinking water supply of the nation's capital \citep{AmericanRivers2026}. Average consumptive water use is approximately 4~MGD, but peak-day use reaches $\sim$15~MGD, an aggregate peaking factor of 3.75. Though data centres account for only 1\% of total withdrawals in the Washington Metropolitan Area, they represent 9\% of annual consumptive use and up to 12\% during summer, when competing outdoor demands are highest and river flows are lowest \citep{ICPRB2026}. Yet the combined delivery capacity of the three utilities serving the region (Fairfax Water, WSSC Water, Washington Aqueduct) exceeds $\sim$800~MGD, yielding an indicative WCI $\approx 15/800 = 1.9\%$, well below the 3--8\% range seen in the small-community sites of Table~\ref{tab:wci_input}. The contrast is striking: the world's densest data-centre cluster imposes modest community-scale burden precisely because it is hosted by a large metropolitan system. This confirms the core WCI insight that burden is a function of facility demand relative to host capacity, not of absolute demand alone. It also suggests that EPRI's projected seven additional states exceeding 20\% data-centre electricity share by 2030 \citep{EPRI2026} may create new small-community burden hotspots as development disperses into regions with smaller utility systems. Under the ICPRB's baseline scenario, consumptive use in the Potomac basin is projected to reach $\sim$22~MGD average and over 80~MGD peak by 2050 \citep{ICPRB2026}, underscoring that even large metropolitan systems face growing cumulative pressure.

Two scope boundaries apply. First, WCI is a utility-impact index, not a hydrologic scarcity model; watershed-scale depletion, groundwater drawdown, and environmental flow impacts require dedicated analysis (see \citep{Amanambu_HSI}). WCI answers a tighter question: can the host utility physically serve the facility on peak day? Second, no standardised dataset links facility demand, PUE, source mix, and utility capacity; the analysis relies on public records, environmental reports, and permit filings. Until comprehensive measurement exists, site-level assessments built on transparent scenarios provide the most defensible basis for siting and permit conditions.

\FloatBarrier

\begin{figure}[H]
\centering
\includegraphics[width=\textwidth]{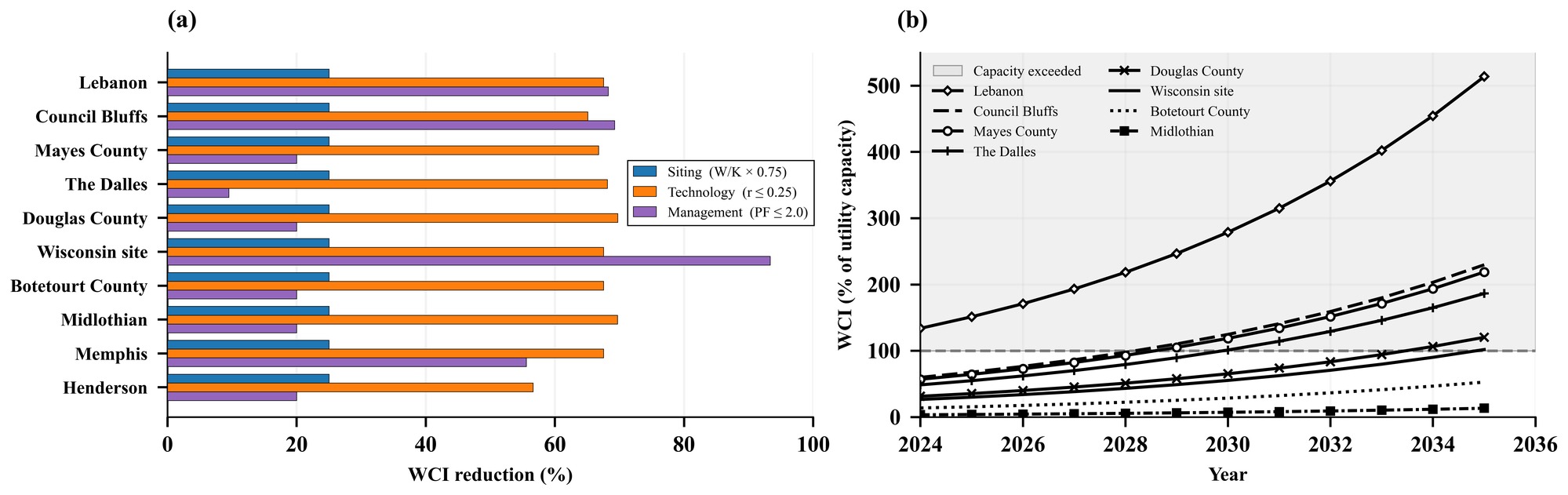}
\caption{\textbf{Policy lever sensitivity and projected WCI growth for ten US data-centre locations.} (a)~Policy lever sensitivity, computed site by site by recalculating WCI with each of the three levers applied in isolation. \textit{Siting} caps demand scale at $W/K' = 0.75 \times W/K$, \textit{Technology} caps the consumptive ratio at $r' = \min(r, 0.25)$, and \textit{Management} caps the peaking factor at $\mathrm{PF}' = \min(\mathrm{PF}, 2.0)$. Bars show the absolute WCI reduction (percentage points of host-utility capacity) delivered by each lever, revealing which factor carries the most leverage at each site: siting returns a flat 25\% everywhere (it scales with $W$); management dominates where PF is large (Lebanon, Council Bluffs, Wisconsin); technology dominates where the consumptive ratio is already high (Douglas County, Mayes County, The Dalles, and the warm-humid small sites). (b)~Projected WCI trajectories from 2024 to 2035 under $g = 0.13$, the lower-bound energy-growth scenario from \citet{Shehabi2024}, with host-utility capacity $K$ held constant. Memphis and Henderson are omitted from panel (b) because their trajectories remain below 5.2\% of host capacity across the entire projection window (maximum 5.15\% and 0.60\% respectively in 2035) and would compress the plotted range of the other eight sites without changing the story. The gray band marks the region above the 100\% capacity threshold; even under this conservative growth rate, five additional sites cross or approach it before 2035, and Lebanon (already above 100\%) continues to diverge.}
\label{fig:wci_growth}
\end{figure}

\subsection{Host Community Typology}
\label{sec:synthesis_typology}

A complementary typology places each community within the broader landscape of hydrologic and infrastructure conditions. The typology classifies hosts along two axes: hydrologic status (water-abundant vs.\ stressed) and infrastructure capacity (adequate vs.\ constrained), with three modifiers capturing important variations.

\begin{enumerate}
    \item \textbf{Type 1: Water-abundant, infrastructure-adequate.} Both source and engineered system have capacity. The binding constraint is regulatory.
    \item \textbf{Type 2: Water-abundant, infrastructure-constrained.} Ample flow but insufficient treatment, distribution, or storage. The Dalles, Oregon exemplifies: the Columbia River delivers enormous flow, but $K = 17$~ML/d gives $\mathrm{WCI} = 0.49$ \citep{OPB2026}. Lebanon, Indiana is extreme ($\mathrm{WCI}_{\mathrm{peak}} = 1.34$). The binding constraint is infrastructure investment and permitting timelines.
    \item \textbf{Type 3: Hydrologically stressed, infrastructure-adequate.} The distribution system has capacity, but the source portfolio is stressed during critical periods. Phoenix exemplifies: Central Arizona Project (CAP) and Salt River Project (SRP) deliver a well-provisioned supply, yet cumulative demand combined with Colorado River shortage declarations pushes HSI above 100\% \citep{Amanambu_HSI}. Infrastructure investment alone cannot solve the problem.
    \item \textbf{Type 4: Dual-stressed.} Both watershed and infrastructure at or near capacity. Douglas County, Georgia approaches this: available water goes negative on the Chattahoochee during low flow, and $\mathrm{WCI} = 0.31$. Small communities in rapidly growing corridors (e.g., Newton County) often fall here \citep{NYT2025}.
\end{enumerate}

\textit{Modifiers.} \textbf{R: Reclaimed water feasible.} The community has wastewater capacity and reuse regulation, shifting the binding constraint from supply to treatment capacity and quality standards (e.g., ``Type 3-R''). \textbf{S: Contested.} The community opposes further development due to accumulated burden or competing demands (e.g., Newton County is Type 4-S; portions of the Great Lakes compete with agricultural irrigation) \citep{AllianceGreatLakes2025}. \textbf{D: Desalination-dependent.} No freshwater source; reliance on energy-intensive desalination couples AI water and energy demand (Gulf states).

\textit{Intervention logic.} Type~1 needs regulatory frameworks. Type~2 needs capital investment (enabling-adaptation pathway is accessible). Type~3 needs technology forcing or siting reallocation. Type~4 is not a candidate for expansion without fundamental system redesign. Modifier~R communities approximate a closed loop. Modifier~D communities require decoupling water availability from desalination energy.

The typology also captures a feedback between thermal burden and water stress: the data heat island effect, with land surface temperatures rising by $\sim$2$^\circ$C (extremes 9$^\circ$C) near clusters \citep{Marinoni2026}, amplifies cooling loads in Type~3 and Type~4 communities. Corporate Water Positive pledges aggregating global replenishment against global consumption obscure this local reality: a facility imposing net burden in a dual-stressed community cannot offset it by funding watershed restoration on another continent.

\subsection{Assumptions and Limitations of the Framework}
\label{sec:synthesis_limitations}

The WCI framework is intentionally scoped as a community-scale utility impact index. It quantifies how strongly a facility's peak consumptive demand presses against the host utility's service capacity. It does not model watershed hydrology, groundwater recharge, environmental flow requirements, or cumulative basin-scale depletion, each of which requires dedicated hydrological analysis. This scope is a strength, not a limitation: by focusing on the utility-capacity dimension, WCI remains computable from publicly available data (facility demand, utility capacity, peaking factor) without requiring the site-specific hydrological modelling that constrains broader scarcity indices to data-rich watersheds.

Because WCI resolves burden at the community level, it exposes a distributional asymmetry: small water systems (99\% of U.S. systems) have the least capacity to absorb hyperscale infrastructure costs \citep{EPA2023}, and national or corporate aggregate Water Positive claims obscure this local reality.

Several assumptions and uncertainties apply. The three-factor decomposition assumes that the consumptive ratio $r$ is constant across seasons and operating conditions; in practice, $r$ varies with ambient temperature, humidity, and cooling system configuration, though the range of 0.70--0.90 is well established \citep{Han2026}. The growth projections (Eq.~\ref{eq:wci_growth}) hold host utility capacity $K$ constant, which is conservative: infrastructure expansions are planned at several sites (e.g., Lebanon, IN, with capacity additions expected by 2031; \citep{CityLebanon2025}), but these expansions are uncertain and would reduce projected WCI values. Conversely, the projections do not account for new facilities co-locating in the same service area, which would increase cumulative demand on the same $K$. The location dominance finding of 4,897 times \citep{Amanambu_HSI} is robust across sensitivity analyses but assumes standardised conditions. Industry growth rates are inherently uncertain; the two scenarios bracket observed rates but do not account for potential acceleration or deceleration. The WCI framework is a first-order analytical tool; the empirical demonstration in Section~\ref{sec:wci_results} relies on operator environmental reports, published utility capacities, and demand estimates from permit filings and public records (Table~\ref{tab:wci_input}; Supplementary Tables~S5--S6), which remain only partially observable for many facilities. PUE values, where available from Google's Environmental Report 2025, provide context for facility energy profiles, but many operators do not disclose site-level PUE. The WCI computation itself does not require PUE, relying instead on directly reported withdrawal volumes.

\begin{figure}[H]
\centering
\includegraphics[width=\textwidth]{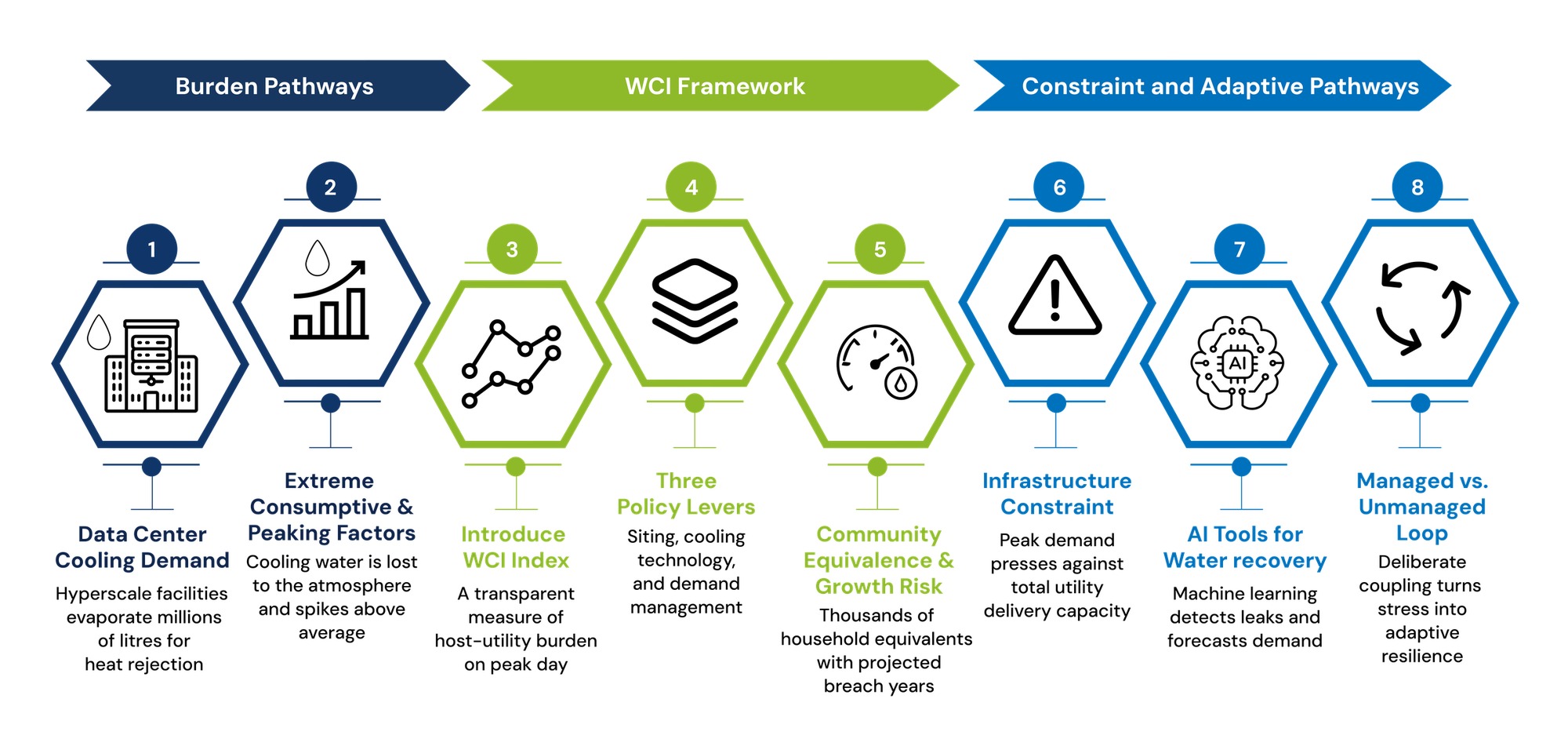}
\caption{Conceptual synthesis of the Water--AI Feedback Loop framework. The figure organises eight linked components spanning: burden pathways (Steps 1--2), which show how data center cooling demand and extreme consumptive and peaking factors stress host water systems; the WCI framework (Steps 3--5), which introduces a community-scale burden index, identifies three actionable policy levers, translates burden into household equivalents, and projects WCI growth and capacity-breach timelines; and constraint and adaptive pathways (Steps 6--8), which show how infrastructure limits reshape AI development and how AI tools can conditionally recover capacity, with the feedback loop's trajectory depending on whether coupling is intentionally managed.}
\label{fig:loops}
\end{figure}

The global pattern of data center clustering in locations chosen for power, connectivity, and land cost rather than water sustainability is visible in the 72\% of Chinese data center capacity sited in water-scarce regions \citep{Jiang2025} and in Ireland's electricity grid carrying 22\% data center load \citep{CSO2025}. A spatially explicit analysis of the S\~{a}o Paulo AI infrastructure cluster illustrates the feedback loop in a hydropower-dependent economy: an estimated annual water footprint of 16.1 million m$^3$, of which over 46\% is indirect water consumed through hydroelectric generation, creating a direct coupling between data centre demand and the hydrological systems that power the grid \citep{Liang2026}. Testing the framework in these settings remains a priority. The full Water--AI feedback architecture, linking burden pathways, the WCI framework, and the conditional coupling between constraint and adaptation, is synthesised in Figure~\ref{fig:loops}.

\begin{table}[H]
\centering
\caption{Quantitative synthesis of key indicators across the three pathways of the Water and AI Feedback Loop.}
\label{tab:quant_synthesis}
\scriptsize
\setlength{\tabcolsep}{7pt}
\renewcommand{\arraystretch}{1.15}
\begin{threeparttable}
\begin{tabularx}{\textwidth}{>{\raggedright\arraybackslash}p{3.8cm}
                              @{\hspace{8pt}}>{\raggedright\arraybackslash}p{1.7cm}
                              >{\raggedright\arraybackslash}p{2.4cm}
                              >{\raggedright\arraybackslash}p{1.3cm}
                              >{\raggedright\arraybackslash}X}
\toprule
\textbf{Parameter} & \textbf{Value} & \textbf{Source} & \textbf{Period} & \textbf{Significance} \\
\midrule
\multicolumn{5}{l}{\textit{Burden Pathways}} \\[1pt]
U.S. DC electricity consumption       & 177--192 TWh       & \citet{Shehabi2024}; \citet{EPRI2026}  & 2024       & Baseline for indirect water estimation \\
Projected DC electricity (LBNL)       & 325--580 TWh       & \citet{Shehabi2024}              & 2028       & Growth driver for all water indicators \\
Projected DC electricity (EPRI)       & 380--790 TWh       & \citet{EPRI2026}                 & 2030       & 60\% above EPRI 2024 estimates \\
U.S. total electricity demand         & 4,108--4,244 BkWh  & \citet{EIA_STEO2026}             & 2026--2027 & DC growth $\gg$ grid growth (1--3\%) \\
Potomac basin DC consumptive use      & 15.1 avg / 56.8 peak ML/d & \citet{ICPRB2026}               & 2025       & 9--12\% of regional consumptive use \\
New water capacity required           & 2,638--5,493 ML/d     & \citet{Han2026}                  & 2024--2030 & Comparable to NYC daily supply \\
Infrastructure cost                   & \$7--58 B          & \citet{Han2026}                  & 2024--2030 & Concentrated on small systems \\
DC consumptive ratio                  & 70--90\%           & \citet{Han2026}  & Current    & vs.\ 12\% public supply average \\
DC peaking factor range               & 3 to $>$30         & \citet{Han2026}  & Current    & vs.\ 1.5--2.5 for other users \\
Location dominance ratio              & 4,897$\times$      & \citet{Amanambu_HSI}             & 30-yr      & Location $\gg$ technology effect \\
HSI range (250 MW facility)           & 0.0008--42.78\%    & \citet{Amanambu_HSI}             & Std.       & 4+ orders of magnitude \\
Global DC in water-stress areas       & 43\%               & \citet{SPGlobal2025}             & 2020s      & Rising to 45\% by 2050s \\
China DC in water-scarce regions      & 72\%               & \citet{Jiang2025}                & Projected  & Severe spatial mismatch \\
Global AI Scope 1+2 water             & 4.2--6.6 B m$^3$  & \citet{Li2025}                   & 2027       & Equivalent to 4--6 Denmarks \\
U.S. AI server water footprint        & 731--1,125 M m$^3$ & \citet{Xiao2025}                 & 2024--2030 & State-level concentration \\
\midrule
\multicolumn{5}{l}{\textit{Constraint Pathways (infrastructure)}} \\[1pt]
Power penalty of forced dry cooling   & 25--35\%           & \citet{Amazon2026}               & Current    & Energy cost of technology substitution \\
Water emerging as binding siting constraint & 43\% of DCs in water-stressed areas & \citet{SPGlobal2025}        & 2020s      & Paradigm shift in siting \\
Lebanon, IN infrastructure delay      & Until 2031         & \citet{CityLebanon2025}          & Planned    & 6+ years from construction start \\
Infrastructure investments required   & \$25M--\$400M+     & Various                          & 2024--2026 & Capital cost of capacity expansion \\
\midrule
\multicolumn{5}{l}{\textit{Adaptive Pathways (conditional)}} \\[1pt]
U.S. non-revenue water                & ~19.5\%            & \citet{BluefieldNRW2025}         & Current    & Exceeds total DC demand \\
AI leak detection accuracy            & 94.2\%             & \citet{WaterTwinAI2025}          & 2025 pilot & 15\% loss reduction reported \\
U.S. water infra.\ funding gap        & \$1.3--3.4 T       & \citet{EPA2023}; \citet{ValueOfWater2025} & 20-year & AI can stretch existing capacity \\
LLM water footprint variation         & 0.3--34 mL/query   & \citet{Jegham2025}               & 2025       & 100$\times$ range across models \\
\bottomrule
\end{tabularx}
\end{threeparttable}
\end{table}

\section{Governance Priorities and Research Frontiers}
\label{sec:agenda}

Critical gaps remain at the interfaces between pathways. Priority needs are organised by interface, each linked to governance implications.

\subsection{Burden to Constraint Interface}

The dynamics through which stress translates into constraint remain poorly quantified. Systematic analysis of the temporal overlap between peak cooling demand, low streamflow, grid stress, and community water peaks is needed. Mapping the unmet water footprint of dry-cooled data centers, the water they would consume if available, and assessing its implications for infrastructure planning and water rights allocation represents a second priority. Projecting how shifting drought frequency and intensity under climate change alter HSI values and curtailment probabilities is essential, as preliminary analysis suggests that Phoenix's ten-year curtailment probability could reach 80\% by 2050 \citep{Amanambu_HSI}.

\textit{Governance implication:} Regulatory frameworks designed for users with peaking factors below 2.5 are structurally inadequate for data centers. New permitting categories that explicitly account for consumptive ratio, peaking factor, and drought-period stress are needed.

\subsection{Constraint to Adaptation Interface}

The constraint-to-adaptation dynamics offer the most actionable research opportunities. Quantifying the avoided infrastructure cost when AI-driven efficiency improvements free enough capacity to accommodate data center demand without capital expansion would demonstrate the economic case for coupling the pathways. Developing unified simulation platforms that embed hydrological stress assessment, WCI decomposition, and AI optimisation simultaneously would bridge the ``river'' and ``pipe'' perspectives \citep{Amanambu_HSI,Han2026}. Field trials of workload scheduling that responds to real-time streamflow forecasts and infrastructure capacity signals would apply water-aware computing in practice.

\textit{Governance implication:} Data center water agreements should be structured as adaptive management contracts that couple water access to investment in host community water system intelligence, creating institutional mechanisms for the enabling-adaptation regime.

\subsection{Adaptation to Burden Interface}

New accounting frameworks are needed to close the loop quantitatively. The WCI framework proposed in Section~\ref{sec:synthesis_net} requires empirical validation across facilities served by diverse utility configurations and source portfolios. Because WCI is scoped as a utility impact index rather than a full hydrological model, extending it to include source-side constraints (groundwater depletion, environmental flow exceedance) through complementary indices remains an important research frontier. Formal optimisation of data center and water infrastructure co-location that minimises WCI while maximising computational capacity remains undeveloped. Assessing the potential for AI-optimised wastewater reclamation to reduce potable consumption without shifting burden elsewhere would quantify one of the most direct adaptive pathways.

\textit{Governance implication:} Water Positive pledges should be supplemented by community-level WCI reporting that quantifies peak-day capacity consumption and local household equivalence, distinguishing local utility burden relief from distant replenishment.

\subsection{Integrated Siting and Community Preparedness}

The evidence converges on a clear governance need: an integrated siting framework that evaluates hydrological capacity, infrastructure adequacy, community resilience, and adaptive potential simultaneously, before a facility is permitted rather than after stress appears. Current siting decisions are driven mostly by power availability, fibre connectivity, and land cost, with water considered, if at all, as a secondary operational input \citep{Siddik2021,Kane2025}. The result is systematic misallocation: data centers cluster in locations that optimise for electricity and latency while externalising water costs onto host communities that had no seat at the siting table. These externalities extend beyond water systems: preliminary econometric evidence from Loudoun County, Virginia, suggests that houses within 0.5~miles of a data center sell for 2.8\% less than comparable properties 0.5 to 1.5~miles away, with aggregate seller losses estimated at 8.8 to 17.5 million USD \citep{Rubinovitz2026}. Rising residential water-sewer rates in host communities add a 
distributional equity dimension to the siting problem: in Newton 
County, Georgia, water costs have soared since data centre 
construction began, with rates projected to increase by 33\% over 
two years \citep{NYT2025}.

Several recent frameworks point toward a more integrated approach. The University of Michigan's data center guidebook for local governments provides a structured assessment template that communities can use to evaluate proposals against their water system capacity, fiscal resilience, and development priorities \citep{UMichigan2026}. The Brookings Institution recommends regional water-aware siting protocols that require developers to demonstrate infrastructure neutrality, meaning that the facility adds at least as much capacity as it consumes, before receiving water service commitments \citep{Kane2025}. The WRI framework identifies seven dimensions of community impact and recommends that local leaders require cumulative impact assessments rather than project-by-project review \citep{Walker2026}. These tools, combined with the WCI framework and host community typology proposed in Section~\ref{sec:synthesis}, provide the analytical foundation for a siting regime that treats water as a first-order constraint rather than an afterthought.

\textit{Governance implication:} State agencies should require water impact assessments as a condition of data center permitting, modelled on the environmental impact assessment process but specifically calibrated to the peak-day, high-consumptive-ratio demand profile that distinguishes data centers from all other public water users.

\subsection{Cross-Cutting Needs}

Per-facility water withdrawal, consumption, and peak demand data 
remain largely proprietary. The pattern of non-disclosure documented 
in U.S. court filings, where operators and municipalities have 
classified basic water use data as trade secrets or confidential 
information, undermines the democratic water governance on which 
effective regulation depends \citep{PostCourier2024,VirginiaMercury2024}. Standardised, audited reporting that includes peak WUE alongside annual WUE is essential for both research and regulation \citep{Han2026}, but the index itself requires reform: WUE should be supplemented by community-contextual indi, such as the WCI index proposed here, that anchor efficiency to host-utility capacity and distinguish freshwater from recycled sources (see also \citep{BarnettItzhaki2026}). The framework must be tested in data center markets with different hydrological baselines, infrastructure conditions, and regulatory regimes, including the EU, China, India, Southeast Asia, and the Middle East.

Communities considering data center proposals need standardised assessment tools that integrate hydrological capacity, infrastructure cost, and peak demand impact. Emerging guidebooks for local governments \citep{UMichigan2026} and siting frameworks \citep{Kane2025} represent initial steps, but these tools require validation against the diverse community typologies identified in Section~\ref{sec:synthesis_typology}. Federal and state agencies should develop model water service agreements and infrastructure financing mechanisms calibrated to the distinctive demand profile of data centers \citep{Walker2026,EPA2024afford}.

\section{Conclusion}
\label{sec:conclusion}

Data centers burden water systems, water constraints reshape AI development, and AI tools can conditionally recover water system capacity. These three dynamics form a feedback loop whose trajectory depends on whether it is intentionally managed.

Three findings emerge from this review. First, AI water stress is a community-scale utility burden problem, not a national volumetric problem. The 0.6 to 1.1\% of national withdrawals masks local stress indices exceeding 100\%, peaking factors of 6 to 30, and infrastructure costs of 7 to 58 billion USD concentrated on communities already facing a 1.3 to 3.4 trillion USD funding gap. Second, water infrastructure is not a passive service but an active constraint: capacity bottlenecks in treatment plants, transmission mains, and storage systems are reshaping AI's geography, architecture, and deployment timeline through infrastructure timelines projected at six years or longer, forced technology substitution, and emerging regulatory conditions on infrastructure access. Third, AI tools that could in principle close the loop, including leak detection, demand forecasting, digital twins, and water-aware scheduling, are technically maturing but remain largely undeployed in the communities most stressed by data center demand, and no peer-reviewed study has demonstrated that AI-driven water savings offset co-located data center burden at operational scale. Their potential depends on local matching of where, when, and at what scale the savings appear.

Two analytical constructs put this argument into practice. The Water Consumption Impact (WCI) index provides a community-scale index for quantifying how strongly a data center's peak consumptive demand presses against the service capacity of its host utility. WCI is intentionally scoped as a utility impact index, not a full hydrologic scarcity model, making it computable from publicly available data and directly actionable for community planning. Its three-factor decomposition gives stakeholders three things immediately: how big the site is relative to the utility ($W/K$), how much water is effectively lost ($r$), and how extreme the peak is ($\mathrm{PF}$), each mapping to a distinct policy lever of siting, cooling technology, and demand management respectively. Application to ten US locations demonstrates that utility burden spans nearly three orders of magnitude: Lebanon, Indiana ($\mathrm{WCI} = 1.34$) already exceeds host capacity, while Henderson, Nevada ($\mathrm{WCI} = 0.002$) imposes negligible burden. Growth projections at observed industry rates show five additional sites breaching the capacity threshold within eleven years. Community equivalence analysis translates these abstract ratios into tangible terms: a single facility can consume water equivalent to thousands of households. The host community typology provides a structured basis for comparing contexts and selecting appropriate interventions. Together with the emerging integrated siting frameworks now being developed by research institutions and policy organisations \citep{UMichigan2026,Kane2025,Walker2026}, these constructs provide the analytical architecture for a water-first approach to AI infrastructure planning, one that evaluates hydrological, infrastructure, and community capacity before siting decisions are made rather than after burdens have materialised.

Closing this feedback loop requires breaking down the disciplinary silos that have kept the AI and water problem fragmented. The hydrologist asking ``Can the river basin handle it?'' and the infrastructure engineer asking ``Can the pipes handle it?'' and the computer scientist asking ``How do we make AI more water-efficient?'' are all studying the same coupled system from different vantage points. The feedback loop framework, the WCI index, and the host community typology provide the conceptual architecture for integrating these perspectives.

The question is no longer whether AI and water systems will co-evolve; they already are. The question is whether the coupling will be managed before local burdens exceed infrastructure and hydrological capacity. Answering it requires connecting what are currently three separate disciplinary conversations, about AI's hydrological burden, infrastructure constraints on AI, and AI's capacity to strengthen water systems, into one.

\section*{Data and code availability}
Data and codes will be made available upon publication.

\section*{Author contributions}
B.A.A. conceptualised the study, developed the methodology, wrote the software, performed the formal analysis, curated the data, conducted the investigation, wrote the original draft, and created the visualisations. A.C.A. conceptualised the study, developed the methodology, supervised the research, reviewed and edited the manuscript, administered the project, and acquired funding. J.M.F. contributed to the methodology and reviewed and edited the manuscript. S.R. contributed to the methodology and reviewed and edited the manuscript. All authors read and approved the final manuscript.

\section*{Competing interests}
All authors declare no financial or non-financial competing interests.

\clearpage

{\LARGE\bfseries\noindent Supplementary Information}
\vspace{1em}

\setcounter{table}{0}
\renewcommand{\thetable}{S\arabic{table}}
\setcounter{figure}{0}
\renewcommand{\thefigure}{S\arabic{figure}}
\setcounter{section}{0}
\renewcommand{\thesection}{S\arabic{section}}
\setcounter{equation}{0}
\renewcommand{\theequation}{S\arabic{equation}}

\section{Constraint Pathway: Source Audit}
\label{sec:supp_constraint_audit}

The constraint pathway (Section~4 of the main text) draws on a broader range of source types than the burden or adaptive pathways, because peer-reviewed research on water infrastructure as an active constraint on AI development remains sparse. Table~\ref{tab:constraint_audit} maps every major claim in the constraint section to its primary source, source type, and evidence status, making visible which claims rest on authoritative public records, which are supported by operator disclosures, and which draw on contextual reporting.

Table~\ref{tab:constraint_audit} reveals a characteristic pattern: the infrastructure constraint mechanisms (capacity bottlenecks, technology forcing, regulatory conditions) rest on strong evidentiary foundations (government records and peer-reviewed analysis), while the institutional mediators (social licence, economic, transboundary) draw more heavily on advocacy reports and journalism. This pattern reflects the maturity of research attention rather than the strength of the phenomena themselves.

\begin{table}[H]
\centering
\caption{Constraint pathway source audit. Each major claim is mapped to its primary source, source type, and whether the claim is directly documented or inferred from available records.}
\label{tab:constraint_audit}
\scriptsize
\setlength{\tabcolsep}{3pt}
\renewcommand{\arraystretch}{1.05}
\begin{tabularx}{\textwidth}{>{\raggedright\arraybackslash\hspace{0pt}}p{1.7cm} >{\raggedright\arraybackslash\hspace{0pt}}p{1.55cm} >{\raggedright\arraybackslash\hspace{0pt}}X >{\raggedright\arraybackslash\hspace{0pt}}p{2cm} >{\raggedright\arraybackslash\hspace{0pt}}p{1.3cm} >{\raggedright\arraybackslash\hspace{0pt}}p{1.6cm}}
\toprule
\textbf{Mechanism} & \textbf{Case} & \textbf{Key Claim} & \textbf{Primary Source} & \textbf{Source Type} & \textbf{Evidence Status} \\
\midrule
\multicolumn{6}{l}{\textit{Hydrological limits (treated in burden pathway, Section~3.1)}} \\
Hydrologic & Phoenix, AZ & HSI exceeds 100\% under cumulative demand & Amanambu et al.\ (2026) & Peer-reviewed & Directly modelled \\
Hydrologic & The Dalles, OR & Reservoir on federal land limits expansion & OPB (2026) & Journalism & Reported; corroborated \\
Hydrologic & China & 72\% of DC capacity in water-scarce regions & Jiang et al.\ (2025) & Peer-reviewed & Directly quantified \\
\midrule
\multicolumn{6}{l}{\textit{Infrastructure Capacity Bottlenecks}} \\
Infrastructure & Newton Co., GA & Water request exceeds surplus; project blocked & NYT (2025) & Journalism & Reported \\
Infrastructure & Lebanon, IN & 8 MGD not available until 2031; multi-year capacity expansion & City of Lebanon (2025) & Government & Directly documented \\
Infrastructure & Port Washington, WI & 1.2 MGD request; \$100M+ upgrade & Spectrum News (2025) & Journalism & Reported \\
Infrastructure & Wisconsin (Microsoft) & 8.4M gal/yr site; peak draw $>30\times$ avg & Wisconsin Public Radio (2025) & Journalism & Reported \\
Infrastructure & 1,455 US utilities & 67M customers; 36\% bonds lack climate disclosure & Lyle et al.\ (2025) & Peer-reviewed & Directly quantified \\
\midrule
\multicolumn{6}{l}{\textit{Forced Technology Substitution}} \\
Technological & Microsoft & Zero-water DC designs as response to constraints & Solomon (2024) & Operator & Self-reported \\
Technological & Southern NV & Evaporative cooling banned in new developments & S\&P Global (2025) & Industry & Documented \\
Technological & Energy penalty & Dry cooling requires 25--35\% more electricity & Amazon (2025) & Operator & Self-reported \\
\midrule
\multicolumn{6}{l}{\textit{Regulatory Conditions on Infrastructure Access}} \\
Regulatory & W.\ Des Moines, IA & Utility requires peak reduction for future projects & WDMWW (2022--25) & Government & Directly documented \\
Regulatory & EU-wide & Regulation 2024/1364 requires WUE disclosure & EU (2024) & Government & Directly documented \\
Regulatory & Susquehanna & Environmental review required for DC withdrawals & SRBC (2025) & Government & Directly documented \\
Regulatory & Thames Water, UK & Options for restricting DC peak water use discussed & Gorey (2025) & Journalism & Reported \\
\midrule
\multicolumn{6}{l}{\textit{Institutional Mediators}} \\
Social licence & Multiple US & \$64B in projects blocked or delayed & FWW (2026) & Advocacy & Aggregated \\
Social licence & 4+ jurisdictions & Formal moratoriums on new DC construction & FWW (2026) & Advocacy & Documented \\
Economic & Leesburg, VA & \$25M water/wastewater upgrade & VA DEQ (2024) & Government & Directly documented \\
Economic & Louisiana & \$400M water infrastructure pledge & Amazon News (2026) & Operator & Self-reported \\
Economic & Lebanon, IN & \$10B+ campus; \$75M+ water infrastructure & City of Lebanon (2025) & Government & Directly documented \\
Economic & Mesa, AZ & Google \$6.08 vs.\ residents \$10.80 per 1000 gal & FWW (2026) & Advocacy & From rate schedules \\
Transboundary & Singapore & Malaysian water dependence; agreements expire 2061 & Multiple & Gov't + reporting & Documented \\
Transboundary & Gulf States & Desalination-dependent; no freshwater source & Multiple & Industry + gov't & Widely documented \\
\bottomrule
\end{tabularx}
\end{table}

\section{Adaptive Pathway: Real-World Deployment Evidence}
\label{sec:supp_adaptive_cases}

The main text notes that the adaptive pathway remains the least empirically grounded pillar of the framework. This section compiles the available real-world deployment evidence to distinguish demonstrated capability from modelled potential.

\begin{table}[H]
\centering
\caption{Real-world deployment evidence for AI-driven water system improvements.}
\label{tab:adaptive_deployments}
\footnotesize
\setlength{\tabcolsep}{3.5pt}
\begin{threeparttable}
\begin{tabularx}{\textwidth}{>{\raggedright\arraybackslash\hspace{0pt}}p{1.6cm} >{\raggedright\arraybackslash\hspace{0pt}}p{1.8cm} >{\raggedright\arraybackslash\hspace{0pt}}X >{\raggedright\arraybackslash\hspace{0pt}}p{1.3cm} >{\raggedright\arraybackslash\hspace{0pt}}p{2.1cm} >{\raggedright\arraybackslash\hspace{0pt}}p{1.8cm}}
\toprule
\textbf{Mechanism} & \textbf{Location / System} & \textbf{Reported Outcome} & \textbf{Scale} & \textbf{Source} & \textbf{Offset Demonstrated?} \\
\midrule
Leak detection & Louisville, KY & 700,000 gal/day identified & City-wide & Trade reporting & No co-located DC \\[3pt]
Leak detection & Thames Water, UK & 15\% leakage reduction in pilot zones & Pilot zones & Trade reporting & No co-located DC \\[3pt]
Leak detection & AI-enhanced digital twin & 94.2\% detection; 15\% loss reduction & Pilot & WaterTwin-AI (2025) & No co-located DC \\[3pt]
Leak detection (GNN) & Multiple networks & 95--99\% detection accuracy & Lab + pilot & Z\'{u}\~{n}iga-Uribe et al. (2026) & No co-located DC \\[3pt]
Reclaimed water & 24 AWS sites globally & 530+ million gallons recycled & Operational & Amazon (2025) & Partial: reduces potable demand \\[3pt]
Digital twins & 8--9 global deployments & 30\% aeration energy reduction; 16.7\% pumping savings & Full-scale & Ghorbani Bam et al. (2025) & No co-located DC \\[3pt]
Water-aware scheduling & WATCH algorithm & 20\% water reduction modelled & Simulated & Islam et al. (2015) & Modelled only \\[3pt]
Flood forecasting & Google, 80+ countries & Real-time warnings & Operational & Trade reporting & Not applicable \\
\bottomrule
\end{tabularx}
\begin{tablenotes}
\footnotesize
\item The final column assesses whether the deployment has been shown to offset water burden from a co-located data center. In no case has this offset been demonstrated at operational scale within a host community, reinforcing the main text's characterisation of the adaptive pathway as conditional potential rather than demonstrated closure.
\end{tablenotes}
\end{threeparttable}
\end{table}

\section{Full Source Classification by Pathway}
\label{sec:supp_sources}

Table~\ref{tab:full_sources} provides the complete classification of all sources cited in the main text, organised by pathway. Sources that contribute to multiple pathways are listed under their primary pathway with cross-pathway contributions noted.

{\footnotesize
\begin{longtable}{>{\raggedright\arraybackslash\hspace{0pt}}p{3.2cm} >{\raggedright\arraybackslash\hspace{0pt}}p{5.8cm} >{\raggedright\arraybackslash\hspace{0pt}}p{2.2cm} >{\raggedright\arraybackslash\hspace{0pt}}p{3.8cm}}
\caption{Complete source classification by pathway.}
\label{tab:full_sources} \\
\toprule
\textbf{Citation Key} & \textbf{Primary Contribution} & \textbf{Pathway} & \textbf{Cross-pathway Role} \\
\midrule
\endfirsthead
\multicolumn{4}{l}{\textit{Table~\ref{tab:full_sources} continued}} \\
\toprule
\textbf{Citation Key} & \textbf{Primary Contribution} & \textbf{Pathway} & \textbf{Cross-pathway Role} \\
\midrule
\endhead
\bottomrule
\endfoot
\multicolumn{4}{l}{\textit{Burden Pathway (34 sources)}} \\[3pt]
Li et al. (2025) & Water footprint of AI models & Burden & Cross-cutting volumetrics \\
Mytton (2021) & DC water consumption review & Burden & --- \\
Siddik et al. (2021) & US DC environmental footprint & Burden & Cross-cutting siting \\
Siddik et al. (2024) & Spatiotemporal water/carbon footprints & Burden & --- \\
de Vries and Gao (2026) & Global DC water/carbon footprints & Burden & --- \\
Xiao et al. (2025) & AI server net-zero pathways & Burden & --- \\
Herrera (2025) & Scenario-based water footprint forecasting & Burden & --- \\
Farf\'{a}n and Lohrmann (2023) & European digital services footprint & Burden & --- \\
Lei and Masanet (2022) & PUE/WUE estimation methodology & Burden & --- \\
Lei et al. (2025) & DC workload water use determinants & Burden & --- \\
Karimi et al. (2022) & Water--energy tradeoffs in hot-arid DC & Burden & --- \\
Amanambu et al. (2026) & HSI framework; location dominance & Burden & Cross-cutting framework \\
Jiang et al. (2025) & Chinese DC water footprint & Burden & Constraint (spatial mismatch) \\
Han et al. (2026) & DC impact on public water systems & Burden & Cross-cutting framework \\
Alkrush et al. (2024) & DC cooling energy review & Burden & --- \\
Jegham et al. (2025) & LLM inference footprint benchmarking & Burden & --- \\
FlexCoolDC (Gnibga et al., 2024) & Cooling flexibility trade-offs & Burden & --- \\
Falk et al. (2025) & AI resource cost (material intensity) & Burden & --- \\
Shehabi et al. (2024) & US DC energy usage report & Burden & --- \\
ICPRB (Ahmed et al., 2025) & Washington metro water supply study & Burden & Constraint \\
Medalie et al. (2025) & US water use statistics & Burden & --- \\
Privette et al. (2026) & DC water footprint transparency & Burden & Cross-cutting governance \\
Sickinger et al. (2018) & NREL hybrid cooling trial & Burden & Adaptive (technology) \\
Han et al. (2024) & DC public health impact & Burden & --- \\
Tao and Gao (2025) & DC health research frontier & Burden & --- \\
SPGlobal (2025) & Global DC water stress analysis & Burden & Constraint \\
CSO (2025) & Ireland DC electricity consumption & Burden & --- \\
Liemberger and Wyatt (2019) & Global non-revenue water quantification & Burden & --- \\
McKinsey (2025) & US DC economic analysis & Burden & Constraint (economic) \\
JLL (2025) & NA DC market report & Burden & Constraint (economic) \\
RolandBerger (2025) & DC water demand analysis & Burden & --- \\
Apple (2025) & Corporate water strategy & Burden & --- \\
Barnett-Itzhaki (2026) & AI water footprint; digital water sobriety & Burden & Cross-cutting governance \\
Liang (2026) & S\~{a}o Paulo AI water footprint; hydropower nexus & Burden & --- \\[6pt]
\multicolumn{4}{l}{\textit{Constraint Pathway (31 sources)}} \\[3pt]
NYT (2025) & Newton County project rejection & Constraint & --- \\
OPB (2026) & The Dalles water expansion & Constraint & --- \\
SpectrumNews (2025) & Port Washington infrastructure & Constraint & --- \\
CityLebanon (2025) & Lebanon water agreement & Constraint & --- \\
Solomon (2024) & Zero-water DC design & Constraint & --- \\
Amazon (2025) & AI water use FAQ & Constraint & --- \\
Amazon News (2026) & Louisiana DC investment & Constraint & --- \\
Meta (2026) & Lebanon DC announcement & Constraint & --- \\
Nakagawa (2023) & Water Positive journey & Constraint & --- \\
MetaLimnoTech (2025) & Volumetric water benefits & Constraint & --- \\
VirginaDEQ (2024) & Leesburg settlement agreement & Constraint & --- \\
Post and Courier (2024) & Dorchester County FOIA case & Constraint & --- \\
WPR (2025) & Milwaukee Riverkeeper FOIA lawsuit & Constraint & --- \\
Virginia Mercury (2024) & NDA investigation; DC non-disclosure & Constraint & --- \\
Gorey (2025) & AI land/water impacts & Constraint & --- \\
EU (2024) & WUE disclosure regulation & Constraint & --- \\
SRBC (2025) & Basin commission DC FAQ & Constraint & --- \\
FWW (2026) & Advocacy case against DCs & Constraint & --- \\
Lyle et al. (2025) & Utility climate risk index & Constraint & --- \\
Smith (2026) & Community-first AI infrastructure & Constraint & --- \\
Microsoft (2025) & Environmental sustainability report & Constraint & --- \\
Google (2025) & Water stewardship portfolio & Constraint & --- \\
DCD (2023, Uruguay) & Google Uruguay facility reformulation & Constraint & --- \\
Mongabay (2023) & Uruguay drought and DC opposition & Constraint & Burden (volumetric demand) \\
Tironi (2025) & Chile Google DC permit reversal & Constraint & --- \\
DCD (2022, Netherlands) & Netherlands DC moratorium & Constraint & --- \\
CRU Ireland (2025) & Ireland DC moratorium & Constraint & --- \\
Singapore IMDA (2024) & Green DC roadmap; WUE mandate & Constraint & Adaptive (regulation) \\
Germany EnEfG (2023) & Energy Efficiency Act; WUE reporting & Constraint & --- \\
Shah (2026) & Four water insecurity concerns about AI datacenters & Constraint & Cross-cutting governance \\
American Rivers (2026) & Potomac designated most endangered US river & Constraint & Burden \\[6pt]
\multicolumn{4}{l}{\textit{Adaptive Pathway (23 sources)}} \\[3pt]
Hayslep et al. (2025) & Genetic programming leak prediction & Adaptive & --- \\
Taloma et al. (2025) & ML for smart water systems review & Adaptive & --- \\
Jun and Jung (2025) & Deep learning leak detection & Adaptive & --- \\
Fu et al. (2022) & Deep learning in urban water mgmt & Adaptive & --- \\
WaterTwin-AI (2025) & AI-enhanced digital twin platform & Adaptive & --- \\
Kratzert et al. (2019) & ML streamflow prediction & Adaptive & --- \\
Nearing et al. (2021) & Hydrological science and ML & Adaptive & --- \\
Negm et al. (2024) & Deep RL for urban water systems & Adaptive & --- \\
Chen et al. (2018) & CloudHeat waste heat harvesting & Adaptive & --- \\
Yang et al. (2024) & Learning-augmented water infra control & Adaptive & --- \\
Islam et al. (2015) & WATCH water-aware load balancing & Adaptive & --- \\
Farah and Shahrour (2024) & Leak detection bibliometric review & Adaptive & --- \\
Z\'{u}\~{n}iga-Uribe et al. (2026) & AI leak detection systematic review & Adaptive & --- \\
Ghorbani Bam et al. (2025) & Digital twin review (147 studies) & Adaptive & --- \\
Vittori et al. (2025) & GNN for water distribution networks & Adaptive & --- \\
GNN\_WDS (Ashraf et al., 2025) & Physics-informed GNN for WDS & Adaptive & --- \\
Daniel et al. (2023) & Utility digitalisation survey & Adaptive & --- \\
Wu et al. (2025) & SCARF water stress metric & Adaptive & Cross-cutting framework \\
Radovanovic et al. (2022) & Carbon-aware computing & Adaptive & --- \\
Amazon (2025, recycled) & AWS recycled water deployments & Adaptive & --- \\
WaterBriefing (2023) & Thames Water acoustic leak detection & Adaptive & --- \\
Sousa et al. (2026) & Water leakage management review (scientific, industrial, commercial) & Adaptive & --- \\
EPA WRAP 2.0 (2026) & Federal water reuse action plan targeting DC cooling & Adaptive & Cross-cutting governance \\[6pt]
\multicolumn{4}{l}{\textit{Cross-Cutting (9 sources)}} \\[3pt]
EPA (2023) & US drinking water infrastructure needs & Cross-cutting & --- \\
EPA (2024) & Water affordability assessment & Cross-cutting & --- \\
Value of Water (2025) & Infrastructure investment benefits & Cross-cutting & --- \\
Kane (2025) & Brookings AI/water strategy & Cross-cutting & --- \\
Walker and Goldsmith (2026) & WRI community impact analysis & Cross-cutting & --- \\
UMichigan (2026) & DC community guidebook & Cross-cutting & --- \\
Ceres (2025) & Water stewardship benchmark & Cross-cutting & --- \\
AllianceGreatLakes (2025) & Great Lakes DC water needs & Cross-cutting & --- \\
WaterAINexus (2025) & Sustainable DC water principles & Cross-cutting & --- \\
\end{longtable}
}

\begin{table}[H]
\centering
\caption{Distribution of sources across pathways and evidence tiers. Aggregate totals corresponding to the detailed classification in Table~\ref{tab:full_sources}.}
\label{tab:source_counts}
\footnotesize
\setlength{\tabcolsep}{6pt}
\begin{threeparttable}
\begin{tabular}{lrrrrr}
\toprule
\textbf{Pathway} & \textbf{Peer-reviewed} & \makecell{\textbf{Gov't/}\\\textbf{Regulatory}} & \makecell{\textbf{Corporate \&}\\\textbf{Industry}} & \makecell{\textbf{Invest./}\\\textbf{Trade}} & \textbf{Total} \\
\midrule
Burden        & 23 & 4  & 8  & 1 & 36 \\
Constraint    & 8  & 7  & 5  & 8 & 28 \\
Adaptive      & 20 & 2  & 2  & 1 & 25 \\
Cross-cutting & 4  & 1  & 3  & 0 & 8  \\
\midrule
\textbf{Total} & \textbf{55} & \textbf{14} & \textbf{18} & \textbf{10} & \textbf{97}\tnote{*} \\
\bottomrule
\end{tabular}
\begin{tablenotes}
\footnotesize
\item[*] Total includes sources covering international constraint cases (Uruguay, Chile, Netherlands, Ireland), regulatory WUE mandates (Singapore, Germany), information-asymmetry cases (Dorchester County FOIA, Milwaukee Riverkeeper, Virginia NDA investigation), and adaptive pathway deployment evidence (Thames Water). See Section~2.3 of the main text for the review logic and source-inclusion criteria.
\item Four evidence tiers are distinguished: peer-reviewed research (highest weight); government and regulatory documents (authoritative for infrastructure data and regulatory actions); corporate disclosures (self-reported, potentially incomplete); and investigative and trade reporting (case-level context).
\end{tablenotes}
\end{threeparttable}
\end{table}

\section{Water Consumption Impact Framework: Supplementary Equations and Implementation Detail}
\label{sec:supp_wci}

This section provides the supporting equations, step-by-step computational procedures, and implementation detail for the Water Consumption Impact (WCI) framework introduced in Section~6.2 of the main text. WCI is scoped as a \textit{community-scale utility burden metric}: it measures how strongly a data center's peak consumptive demand presses against the service capacity of its host utility. It does not model watershed-scale scarcity, groundwater depletion, or environmental flow requirements, each of which requires dedicated hydrological analysis. This intentional scope makes the metric computable from publicly available data (facility demand, utility capacity, peaking factor) and directly actionable for community planning. The four core equations (Eqs.~1--4: WCI definition, three-factor decomposition, household equivalence, and growth projection) are presented in the main manuscript. The equations below define the intermediate quantities and community-scale metrics in sufficient detail for independent replication.

\subsection{Average daily consumptive use}

The average daily consumptive use $C_{\mathrm{avg}}$ is the volume of water permanently lost to evaporation per day under normal operating conditions:
\begin{equation}
C_{\mathrm{avg}} = r \times W_d
\label{eq:c_avg}
\end{equation}
where $r$ is the consumptive ratio (dimensionless, typically 0.70--0.90 for evaporative cooling systems) and $W_d$ is the average daily withdrawal (ML/d). The consumptive ratio represents the fraction of withdrawn water that is evaporated and not returned to the source watershed. For Google-operated campuses we compute $r$ directly as $r = C/W$ from the site-specific withdrawal and consumption values reported in Google's 2025 Environmental Report (Google, 2025, p.~110; FY2024 data, independently assured by EY). These site-specific ratios span 0.576 (Henderson, NV, which discharges a large share of cooling water back to the SNWA system) to 0.826 (Douglas County, GA, warm-humid climate with high evaporative loss). For non-Google operators (Meta at Lebanon, IN; the Wisconsin hyperscale site operated by Microsoft; xAI at Memphis, TN; and Google Botetourt County, which remains at permitted reservation only), we use $r = 0.77$, the weighted mean reported in Han et al.\ (2026), Appendix~A.5, consistent with the 0.70--0.90 range. Where site-specific WUE data are available, $W_d$ can be computed as $W_d = \mathrm{WUE} \times E_{\mathrm{IT}}$, where $E_{\mathrm{IT}}$ is the IT energy load and WUE is defined per the Green Grid standard as annual site water usage divided by IT equipment energy (L/kWh). PUE values from Google (2025) (p.~110) are reported for Google campuses (Council Bluffs 1.09, The Dalles 1.08, Mayes County 1.11, Douglas County 1.09, Henderson 1.09) and are used in verifying energy-to-water conversions but do not enter the WUE-based withdrawal computation directly.

\subsection{Peak day consumptive use}

The peak day consumptive use $C_{\mathrm{peak}}$ amplifies the average daily consumption by the peaking factor:
\begin{equation}
C_{\mathrm{peak}} = C_{\mathrm{avg}} \times \mathrm{PF}
\label{eq:c_peak}
\end{equation}
where $\mathrm{PF}$ is the peaking factor, the ratio of peak-day to average-day water demand. Peaking factors for data centers are substantially higher than for conventional municipal users (typically 1.5--2.5): published site-specific values span 2.21 to approximately 30 (the Wisconsin hyperscale Microsoft site, reflecting highly intermittent cooling in shared-capacity arrangements; primary records from \citep{WisPubRadio2025}, also discussed in Han et al.\ (2026), \S3.1.2), with Lebanon, IN at PF $=6.3$ (Han et al.\ 2026, Appendix~D, an analytical recomputation), Council Bluffs, IA at PF $=6.5$ (Han et al.\ 2026, \S3.1.2, applying a 1.5 regulatory safety factor to the 4.30 measured monthly peak in West Des Moines Water Works 2022--2025 reports), and the Prince William Water service area in Northern Virginia at a measured daily PF of 10 (Han et al.\ 2026). The peaking factor captures the combined effect of seasonal temperature variation (which drives cooling demand), operational ramp-up patterns, and the extreme thermal loads associated with AI training workloads. For warm-climate sites where no site-specific PF has been reported in peer-reviewed or utility-document sources (Mayes County, OK; Douglas County, GA; Botetourt County, VA; Midlothian, TX; Henderson, NV), we apply a default PF $= 2.5$; for Memphis, TN (xAI Colossus) we apply PF $= 4.5$ as a placeholder reflecting the extreme thermal load profile of a large AI training cluster. All site-specific PF assignments are documented in footnotes to main-text Table~5 (WCI input data).

\subsection{Community water share}

The data center's share of host community water consumption contextualises the facility's demand relative to the residential population it serves:
\begin{equation}
\mathrm{DC}_{\mathrm{share}} = \frac{C_{\mathrm{avg}}}{\mathrm{pop} \times \mathrm{pc}} \times 100\%
\label{eq:dc_share}
\end{equation}
where $\mathrm{pop}$ is the host community population and $\mathrm{pc}$ is the per-capita daily water consumption (ML/person/day). For U.S. communities, $\mathrm{pc} = 0.000303$~ML/person/day (80~gallons/person/day residential use, from USGS 2015 Circular~1441; \citep{USGS2018}). This choice is internally consistent with the household denominator used for the household-equivalence metric ($H_{\mathrm{avg}} = 0.4146$~ML/household/year, equivalent to roughly 300 gallons/household/day with an average U.S. household of 2.63 persons). DC share therefore expresses the data center's consumptive demand as a percentage of the total residential water use in the host community. This metric complements WCI by expressing the burden in population-equivalent terms rather than infrastructure-capacity terms, making it accessible to non-technical stakeholders in community planning processes.

\subsection{Step-by-step computational procedure}

The WCI computation for a given site proceeds as follows:

\begin{enumerate}
\item \textbf{Assemble input data.} For each site, obtain: average daily withdrawal $W_d$ (ML/d), consumptive ratio $r$ (dimensionless), peaking factor PF (dimensionless), host utility maximum deliverable capacity $K$ (ML/d), host community population (persons), and per capita water consumption $\mathrm{pc}$ (ML/person/day). Where available, record site-specific PUE from operator environmental reports for contextual documentation, though PUE does not enter the WCI computation directly.

\item \textbf{Compute average daily consumptive use.} Apply Eq.~\ref{eq:c_avg}: $C_{\mathrm{avg}} = r \times W_d$.

\item \textbf{Compute peak day consumptive use.} Apply Eq.~\ref{eq:c_peak}: $C_{\mathrm{peak}} = C_{\mathrm{avg}} \times \mathrm{PF}$.

\item \textbf{Compute WCI.} Apply main text Eq.~1: $\mathrm{WCI} = C_{\mathrm{peak}} / K$. Values exceeding 1.0 indicate that peak cooling demand exceeds total host utility capacity.

\item \textbf{Decompose WCI into three factors.} Apply main text Eq.~2: $\mathrm{WCI} = (W_d/K) \times r \times \mathrm{PF}$. Identify the dominant factor by comparing the relative magnitude of each component across sites.

\item \textbf{Compute household equivalence.} Apply main text Eq.~3: $\mathrm{HH}_{\mathrm{equiv}} = (C_{\mathrm{avg}} \times 365.25) / H_{\mathrm{avg}}$, where $H_{\mathrm{avg}} = 0.4146$~ML/household/year is the U.S. average household water consumption ($\approx$300~gallons/household/day; \citep{EPA_WaterSense2024}).

\item \textbf{Compute community water share.} Apply Eq.~\ref{eq:dc_share}: $\mathrm{DC}_{\mathrm{share}} = C_{\mathrm{avg}} / (\mathrm{pop} \times \mathrm{pc}) \times 100\%$.

\item \textbf{Project WCI forward.} Apply main text Eq.~4: $\mathrm{WCI}(t) = \mathrm{WCI}_0 \times (1+g)^t$ for each year $t = 1, 2, \ldots, 12$ (2024--2035) using a compound annual growth rate $g = 0.13$, the lower bound of projected U.S.\ data-centre energy consumption growth from Shehabi et al.\ (2024). Identify the breach year as the first $t$ where $\mathrm{WCI}(t) \geq 1.0$. 

\end{enumerate}

\subsection{Data sources and implementation notes}

The framework is designed for use under current data limitations. In practice:
\begin{itemize}
\item $W_d$ may come from utility billing records, service agreements, operator environmental reports, or scenario-based demand estimates. For the six Google campuses (Council Bluffs, Mayes Co., The Dalles, Douglas Co., Midlothian, Henderson), we use the site-specific withdrawal values reported in Google's 2025 Environmental Report (Google, 2025), p.~110, which reflect FY2024 data independently assured by EY. For Meta (Lebanon, IN), we use 1.27~MGD ($=4.81$~ML/d), which is the average intake corresponding to the analytical recomputation of Lebanon in Han et al.\ (2026), Appendix~D, rather than a directly reported present-day withdrawal (the measured Project Domino agreement \citep{CityLebanon2025} specifies a phased buildout reaching 8~MGD by 2031). For xAI (Memphis, TN), we use 1~MGD ($=3.79$~ML/d) from the GEAA AI-Water Fact Sheet (2025). For the Wisconsin hyperscale site operated by Microsoft (public records obtained by Wisconsin Public Radio, September~2025; \citep{WisPubRadio2025}), we use the reported 8.4~million~gallons-per-year operating envelope, corresponding to 0.023~MGD ($=0.087$~ML/d) on an averaged basis. For Botetourt Co., VA, we use the 2~MGD permitted reservation ($=7.57$~ML/d) as the current-phase proxy.
\item $r$ is computed site-by-site as $r = C/W$ from Google (2025), p.~110 (FY2024 data, EY-assured) for the six Google campuses, yielding values that range from 0.576 (Henderson, NV) to 0.826 (Douglas Co., GA); Midlothian, TX yields $r = 0.825$. For non-Google operators and for Botetourt Co. (still at permitted-reservation stage), $r = 0.77$ is assigned as the weighted mean from Han et al.\ (2026), Appendix~A.5, consistent with the 0.70--0.90 range for evaporative cooling systems.
\item PF is site-specific where reported and default where not. Site-specific values: Lebanon, IN PF $=6.3$ (Han et al.\ 2026, Appendix~D, an analytical recomputation assuming 0.20~L/kWh cooling intensity and 1~GW IT load); Council Bluffs, IA PF $=6.5$ (Han et al.\ 2026, \S3.1.2, applying a 1.5 regulatory safety factor to the 4.30 measured monthly peak reported by West Des Moines Water Works, 2022--2025); The Dalles, OR PF $=2.21$ (\citep{ConsorDalles2024}, Consor Water System Master Plan Update for The Dalles, OR, 2024); and the Wisconsin hyperscale site (Microsoft) PF $\approx 30$ (reported to exceed 30 by \citep{WisPubRadio2025}, and adopted in Han et al.\ 2026, \S3.1.2, to illustrate extreme intermittency of cooling draw in shared-capacity arrangements). For Memphis, TN we apply PF $=4.5$ as a placeholder for the thermal load profile of a large AI training cluster. For the remaining warm-climate sites with no published peaking factor (Mayes Co., Douglas Co., Botetourt Co., Midlothian, Henderson) we apply a default PF $=2.5$.
\item $K$ is from state environmental agency permits, utility records, and published water supply plans (sources documented per site in Supplementary Table~S5).
\item Population data are from U.S.\ Census Bureau 2020 estimates.
\item Per capita consumption ($\mathrm{pc} = 80$~gallons/person/day, or $0.000303$~ML/person/day) reflects residential indoor-plus-outdoor use from \citet{USGS2018} (USGS Circular 1441, 2015 water-use estimates). The household denominator $H_{\mathrm{avg}} = 0.4146$~ML/household/year ($\approx$300~gallons/household/day) used for the household-equivalence metric (main text Eq.~3) is from EPA WaterSense \citep{EPA_WaterSense2024}.
\end{itemize}

The equations above do not eliminate uncertainty. Rather, they move uncertainty into explicit and inspectable assumptions while providing a reproducible computational procedure. Input data are documented in main-text Table~5 (WCI input data); host utility capacity and computed results are documented in Supplementary Tables~S5--S6 and in the accompanying Excel workbook.

\section{Data and Computed Results for the WCI Framework}
\label{sec:supp_empirical}

This section documents the input data and computed results for the empirical demonstration of the Water Consumption Impact framework presented in Section~6.2 of the main text. All values are drawn from public sources; data provenance is noted in each table. The accompanying Excel workbook provides formula-linked computations for all values reported here.


\begin{table}[H]
\centering
\caption{Host public water system delivery capacity $K$ and community context for each data center location. $K$ is the maximum deliverable capacity (ML/d). Population and per capita consumption are used for household equivalence (main text Eq.~3) and community share (Eq.~S3) computations. Symbol legend: ``---'' $=$ absent from accessible public disclosure; $^\dagger$ $=$ cumulative capacity of multiple treatment plants.}
\label{tab:wci_capacity}
\footnotesize
\setlength{\tabcolsep}{3pt}
\begin{threeparttable}
\begin{tabular}{@{}l >{\raggedright\arraybackslash}p{2.8cm} r r r l@{}}
\toprule
\textbf{Location} & \textbf{Utility} & \textbf{$K$} & \textbf{$K$} & \textbf{Pop.} & \textbf{Source} \\
 & & (MGD) & (ML/d) & & \\
\midrule
Lebanon, IN          & Lebanon Utilities         & 4.6   & 17.41    & 18,000  & Domino Agmt.\tnote{h} \\
Council Bluffs, IA   & Council Bluffs WW         & 30$^\dagger$  & 113.56   & 63,000  & CBWW\tnote{b} \\
Mayes Co., OK        & Grand River Dam Auth.     & 10    & 37.85    & 41,000  & OK DEQ\tnote{g} \\
The Dalles, OR       & City of The Dalles        & 4.5   & 17.03    & 16,000  & \citet{ConsorDalles2024}\tnote{a} \\
Douglas Co., GA      & Douglasville-Douglas WSA  & 8     & 30.28    & 146,000 & GA EPD\tnote{e} \\
Wisconsin site       & Local PWS (WI)            & 2     & 7.57     & 12,000  & WI DNR\tnote{i} \\
Botetourt Co., VA    & Western VA Water Auth.    & 28    & 105.99   & 34,000  & WVWA\tnote{j} \\
Memphis, TN          & MLGW                      & 258   & 976.64   & 630,000 & MLGW\tnote{k} \\
Midlothian, TX       & City of Midlothian        & 36    & 136.27   & 40,000  & TCEQ\tnote{c} \\
Henderson, NV        & SNWA                      & 900$^\dagger$  & 3,406.87 & 320,000 & SNWA\tnote{f} \\
\bottomrule
\end{tabular}
\begin{tablenotes}
\footnotesize
\item[a] \citet{ConsorDalles2024}: Consor Water System Master Plan Update for The Dalles, OR, November 2024, $K = 4.5$~MGD firm supply.
\item[b] Council Bluffs Water Works (CBWW), General Information. $^\dagger$$K = 30$~MGD is the cumulative capacity of two treatment plants: Narrows Water Treatment Plant (20~MGD) and Council Point Water Purification Plant (10~MGD expanded).
\item[c] Texas Commission on Environmental Quality (TCEQ), public water system records; City of Midlothian combined treatment capacity $K=36$~MGD (Tayman Plant 12~MGD + Auger Plant 24~MGD).
\item[e] Georgia Environmental Protection Division, public water system records.
\item[f] Southern Nevada Water Authority (SNWA), Our Regional Water System. $^\dagger$$K = 900$~MGD is the cumulative treatment capacity of two facilities: Alfred Merritt Smith Water Treatment Facility (600~MGD) and River Mountains Water Treatment Facility (300~MGD).
\item[g] Oklahoma Department of Environmental Quality, public water supply permit.
\item[h] Project Domino Water and Wastewater Agreement (Orla LLC/Meta Platforms), IDEM permit IN5206003; $K=4.6$~MGD permitted intake capacity.
\item[i] Wisconsin Department of Natural Resources, public water system records.
\item[j] Western Virginia Water Authority, Carvins Cove Water Treatment Facility; $K=28$~MGD treatment capacity. $W=2$~MGD permitted reservation per Utility Services Funding Agreement.
\item[k] \citet{MLGW2022}: Memphis Light, Gas and Water Drought Management Plan (September 2022); $K=258$~MGD total system capacity.
\item Population data are rounded to the nearest thousand from \citet{USCensus2020} estimates for the host city or county. Per-capita residential water use: 80~gallons/person/day ($=0.000303$~ML/person/day; from \citep{USGS2018}, USGS 2015 Circular~1441, residential indoor and outdoor combined).
\end{tablenotes}
\end{threeparttable}
\end{table}

\begin{table}[H]
\centering
\caption{Computed Water Consumption Impact (WCI) and community equivalence metrics for analysed data center locations. WCI $= C_{\mathrm{peak}}/K$ (Eq.~1). Three-factor decomposition: $W/K$ (demand scale), $r$ (consumptive ratio), PF (peaking factor). $\mathrm{HH}_{\mathrm{equiv}}$ is household equivalents (Eq.~3). $\mathrm{DC}_{\mathrm{share}}$ is data center share of community water demand (Eq.~S3). Dominant factor identifies which of the three factors contributes most to cross-site variation in WCI. Breach-year symbol ``Beyond 2035'' $=$ 13\% CAGR projection does not cross $\mathrm{WCI}=1$ within the 2024--2035 horizon.}
\label{tab:wci_results}
\scriptsize
\setlength{\tabcolsep}{2.5pt}
\begin{threeparttable}
\begin{tabular}{@{}l c c c c c c c l@{}}
\toprule
\textbf{Location} & \textbf{WCI} & \textbf{$W/K$} & \textbf{$r$} & \textbf{PF} & \makecell{\textbf{HH}\\\textbf{equiv}} & \makecell{\textbf{DC}\\\textbf{share (\%)}} & \makecell{\textbf{Breach yr}\\\textbf{(13\% CAGR)}} & \textbf{Dominant} \\
\midrule
Lebanon, IN          & 1.34\tnote{a} & 0.276 & 0.77  & 6.3  & 3,261 & 67.9 & ALREADY       & PF \\
Council Bluffs, IA   & 0.60  & 0.129 & 0.716 & 6.5  & 9,220 & 54.8 & 2029          & PF \\
Mayes Co., OK        & 0.57  & 0.303 & 0.752 & 2.5  & 7,610 & 69.5 & 2029          & $W/K$ \\
The Dalles, OR       & 0.49  & 0.281 & 0.784 & 2.21 & 3,301 & 77.3 & 2030          & $W/K$ \\
Douglas Co., GA      & 0.31  & 0.152 & 0.826 & 2.5  & 3,348 & 8.6  & 2034          & $r$ \\
Wisconsin site       & 0.27  & 0.012 & 0.77  & 30   &    59 & 1.8  & 2035          & PF \\
Botetourt Co., VA    & 0.14  & 0.071 & 0.77  & 2.5  & 5,136 & 56.6 & Beyond 2035   & None (low) \\
Memphis, TN          & 0.013 & 0.004 & 0.77  & 4.5  & 2,568 & 1.5  & Beyond 2035   & None (low) \\
Midlothian, TX       & 0.035 & 0.017 & 0.825 & 2.5  & 1,665 & 15.6 & Beyond 2035   & None (low) \\
Henderson, NV        & 0.002 & 0.001 & 0.576 & 2.5  & 1,893 & 2.2  & Beyond 2035   & None (low) \\
\bottomrule
\end{tabular}
\begin{tablenotes}
\scriptsize
\item[a] WCI $>$ 1.0 indicates that peak-day consumptive demand exceeds the host utility's total deliverable capacity.
\item Breach year is the first year in which projected $\mathrm{WCI}(t) \geq 1.0$ under 13\% CAGR, the lower bound of U.S.\ data-centre energy consumption growth projected by LBNL (Shehabi et al., 2024). ``ALREADY'' indicates $\mathrm{WCI}_0 \geq 1.0$. ``Beyond 2035'' indicates that the 13\% CAGR trajectory does not cross $\mathrm{WCI}=1$ within the 2024--2035 projection horizon.
\item HH equiv computed using $H_{\mathrm{avg}} = 0.4146$~ML/household/year (U.S. average, $\approx$300 gal/day/household from \citep{USGS2018}). DC share computed using per-capita residential consumption of 80~gal/person/day ($=0.000303$~ML/person/day; USGS 2015 Circular~1441) and rounded 2020 Census population. The ``Dominant'' column identifies the factor that drives the WCI at each site: PF (peak amplification), $W/K$ (demand scale relative to utility capacity), or $r$ (consumptive efficiency). Sites marked ``None (low)'' have WCI $<0.15$ and no single factor exerts meaningful pressure.
\end{tablenotes}
\end{threeparttable}
\end{table}


\nocite{Amazon2025}
\nocite{Ceres2025,ConsorDalles2024,FWW2026,GEAA2025,Google2025Stewardship,%
GoogleBotetourtFAQ,Han2024,IADNRWaterSupply,Islam2015,MLGW2022,Microsoft2025,%
Nakagawa2023,SNWA2024,Smith2026,TaoGao2025,USCensus2020,USGS2018,%
WisPubRadio2025,Wu2025}

\bibliographystyle{unsrtnat}
\bibliography{main,supp}

\end{document}